\documentclass[journal]{IEEEtran}
\usepackage{bbm}
\usepackage{caption}
\usepackage{subfigure}
\usepackage{graphicx}
\usepackage{latexsym}
\usepackage{diagbox}
\usepackage{changepage}
\usepackage[fleqn]{amsmath}
\usepackage{amsmath}
\usepackage{amsfonts}
\usepackage{indentfirst}
\usepackage{CJK}
\usepackage{indentfirst}
\usepackage[varg]{txfonts}
\usepackage{stfloats}
\usepackage{multirow}%
\usepackage{booktabs}
\usepackage{color,soul}
\usepackage{epstopdf}
\usepackage{float}
\usepackage{bm}
\usepackage{makecell}
\usepackage{array}
\usepackage{bm}
\usepackage{mathtools}
\usepackage{geometry}
\usepackage[linesnumbered,ruled,commentsnumbered,longend]{algorithm2e}
\usepackage[linesnumbered,ruled,commentsnumbered,longend]{algorithm2e}
\makeatletter

\newcommand{\Rmnum}[1]{\expandafter\@slowromancap\romannumeral #1@}

\makeatother
\makeatletter

\newcommand{\qed}{\nobreak \ifvmode \relax \else
	\ifdim\lastskip<1.5em \hskip-\lastskip
	\hskip1.5em plus0em minus0.5em \fi \nobreak
	\vrule height0.75em width0.5em depth0.25em\fi}

\SetAlgoLongEnd

\begin{document}

\title{Beamforming Design for the Distributed RISs-aided THz Communications with Double-Layer True Time Delays}
\author{Gangcan Sun, Wencai Yan, Wanming Hao, \emph{Member, IEEE}, Chongwen Huang,~\emph{Member, IEEE},  Chau Yuen,~\IEEEmembership{Fellow,~IEEE}
\thanks{G. Sun, W. Yan and W. Hao are with the School of Electrical and Information Engineering, Zhengzhou University, Zhengzhou 450001, China.
 (E-mail: iegcsun@zzu.edu.cn, yanwencai001@163.com, iewmhao@zzu.edu.cn)}
\thanks{C. Huang is with the College of Information Science and Electronic Engineering, Zhejiang University, Hangzhou 310027, China (E-mails: chongwenhuang@zju.edu.cn)}
\thanks{C. Yuen is with the Singapore University of Technology and Design, Singapore 487372, Singapore (E-mail:  yuenchau@sutd.edu.sg)}
}

\maketitle
\begin{abstract}
In this paper, we investigate the reconfigurable intelligent surface (RIS)-aided terahertz (THz) communication system with the sparse radio frequency chains antenna structure at the base station (BS). To overcome the beam split of the BS, different from the conventional single-layer true-time-delay (TTD) scheme, we propose a double-layer TTD scheme that can effectively reduce the number of large-range delay devices, which involve additional insertion loss and amplification circuitry. Next, we analyze the system performance under the proposed double-layer TTD scheme. To relieve the beam split of the RIS, we consider multiple distributed RISs to replace an ultra-large size RIS. Based on this, we formulate an achievable rate maximization problem for the distributed RISs-aided THz communications via jointly optimizing the hybrid analog/digital beamforming, time delays of the double-layer TTD network and reflection coefficients of RISs. Considering the practical hardware limitation, the finite-resolution phase shift, time delay and reflection phase are constrained. To solve the formulated problem, we first design an analog beamforming scheme including optimizing phase shift and time delay based on the RISs' locations. Then, an alternatively optimization algorithm is proposed to obtain the digital beamforming and reflection coefficients based on the minimum mean square error and coordinate update techniques. Finally, simulation results show the effectiveness of the proposed scheme.
\end{abstract}

\begin{IEEEkeywords}
THz communication, double-layer TTD network, beam split, reconfigurable intelligent surface, hybrid beamforming.
\end{IEEEkeywords}

\IEEEpeerreviewmaketitle

\section{Introduction}
To achieve data rates of terabits-per-second (Tb/s), the future sixth-generation (6G) wireless communications are expected to exploit the terahertz (THz) frequency (0.1-10 THz) due to its ultra-wide bandwidth~\cite{ref1}-\cite{ref21}. However, THz signals usually suffer from the severe path loss and poor diffraction~\cite{ref22}~\cite{ref23}, which leads to the limited coverage. Fortunately, the massive multiple-input multiple-output (MIMO) and reconfigurable intelligent surface (RIS) techniques nowadays are developed~\cite{ref24}, and they can be applied for THz communications to form the high-gain directional beams and improve the signals' coverage~\cite{ref25}. Furthermore, the virtual line of sight (LoS) link can be constructed via the deployment of RISs, and thus the serious blockage problem can be effectively solved~\cite{ref26}-~\cite{ref2600}. Therefore, it is promising for the applications of RIS-aided massive MIMO in future THz communications.
\subsection{Related Works}
It is well known that the base station (BS) can deploy a large number of antennas within the limited physical size because of the small wavelength of THz signals. However, the fully digital beamforming technique requires a unique radio frequency (RF) chain connecting to each antenna, and this will lead to huge power consumption and hardware complexity~\cite{ref261}~\cite{ref262}, which is infeasible in practice. To address this, the hybrid analog/digital beamforming technique is developed~\cite{ref27}~\cite{ref28}, where the antennas are connected to a few RF chains via several groups of phase shifters (PSs). It has been proved that a asymptotically optimal performance can be obtained by optimizing the analog/digital beamforming in the narrowband system. However, due to the frequency-independent characteristic of PSs, the beam split will occur for the wideband THz system, leading to the serious performance loss~\cite{ref29}~\cite{ref291}. Meanwhile, the RIS also faces the similar problem because of its frequency-independent reflecting elements~\cite{refD},~\cite{refD_0}.

Currently, there have been several works studying how to mitigate the beam split effect. One straightforward solution combating beam split is to replace all PSs by frequency-dependent true-time-delays (TTDs)~\cite{ref3}, while this will cause huge power consumption and hardware complexity due to the use of large numbers of TTDs. Instead, a limited number of TTDs are inserted between the RF chains and PSs to solve the beam split~\cite{ref23},~\cite{ref4}-\cite{refJ2}, and thus, the traditional one-dimensional analog beamforming is converted into two-dimensional analog beamforming via the joint control of PSs and TTDs. Specifically, a novel THzPrism architecture is designed in~\cite{ref4}, where TTDs are arranged in a serial manner. Since TTDs with the equal number of antennas are utilized, this architecture will cause a high hardware complexity. Consequently, the authors~\cite{ref23} propose two low hardware complexity schemes, including the virtual subarray beamforming scheme and the sparse TTD-based scheme.  The former does not need to add any extra hardware, while its performance is lower than the latter. Similarly, a TTD-based delay-phase precoding is proposed in~\cite{ref5}, and then the authors extend it to the user position by the beam tracking~\cite{refJ2}. However, the TTDs in~\cite{ref23}~\cite{ref5}~\cite{refJ2} are arranged in a parallel manner, and each TTD must be configured independently with the price of supporting a large-range delay~\cite{refA}~\cite{refB}, especially for the large antenna array. In addition, the transceiver RF signal amplification circuit is needed to compensate the loss of the large delay line, which increases the hardware cost. In order to reduce the required delay range, a hybrid TTD architecture is proposed for fast beam training, where the time delays are separately realized by analog TTDs and digital TTDs~\cite{refC}. Nevertheless, the analog phases and delays are fixed such that the number of required analog TTDs is equal to the number of antennas. To realize the high energy efficiency, a novel energy-efficient dynamic-subarray with fixed TTD architecture is developed~\cite{refC1}. However, the improved performance is very limited based on the schemes of~\cite{refC}~\cite{refC1}. Meanwhile, all the above works assume that TTDs can provide delay with high resolution or even infinite resolution, and this is power-hungry and even infeasible in practice.

To overcome the beam split effect with low hardware complexity at the BS, we propose a double-layer TTD scheme. On the one hand, the proposed scheme can solve the maximum delay compensation problem observed in the traditional single-layer TTD scheme. Specifically, implementing a large-range delay TTD requires serious sacrifices in terms of linearity, noise, power and area, which increase the complexity of the design. Besides, the increase of delay range not only reduces the accuracy, but also deteriorates the nonlinear performance of the system due to the use of a large number of active amplifier devices~\cite{refC2}. Thus, minimizing the number of large-range delay TTDs used in the system would significantly reduce the hardware cost. On the other hand, based on the practical hardware limitation, the discrete time delays with different resolutions at each layer TTD network are considered. In this way, the TTDs of each layer only need to compensate the propagation delay across the specific subarray aperture. In addition, the PSs are used to compensate for the residual phase shift of the double-layer TTD network and generate a beam towarding to the target's physical direction.

Besides, to the best of our knowledge, there has not been the related work jointly considering the beam split effect and beamforming design problems in wideband THz RIS communications. Therefore, we extend the double-layer TTD scheme to the wideband THz distributed RISs communications system to cooperatively mitigate the beam split effect. In fact, the beam split effect of distributed RISs with fewer elements is less severe than that of the centralized RIS~\cite{refD}~\cite{refE}.

\subsection{Main Contributions}
In this paper, we investigate the antenna structure design and beamforming optimization in the RIS-aided THz communications, and the main contributions are summarized as follows.
\begin{itemize}
\item[$\bullet$]
To overcome the beam split of the BS with low hardware complexity, we propose a double-layer TTD scheme. Different from the conventional single-layer TTD scheme, the required number of large-range delay TTDs is sharply reduced by bringing an additional small-range delay TTD network, which effectively reduce the hardware cost. We analyze the phase compensation error and normalized array gain under the proposed double-layer TTD scheme. The results show that the proposed scheme can almost obtain the same performance with the conventional single-layer TTD scheme, but the overall hardware cost is effectively decreased.
\end{itemize}
\begin{itemize}
\item[$\bullet$]
Next, we extend the double-layer TTD scheme to the wideband THz distributed RISs communications system to cooperatively mitigate the beam split effect. In fact, the beam gain loss of distributed RISs with fewer elements is less severe than that of the centralized RIS. Then, we formulate a achievable rate maximization problem via jointly optimizing the hybrid analog/digital beamforming, time delays of the double-layer TTD network and reflection coefficients of RISs. Meanwhile, the finite-resolution phase shifter, time delay and reflection phase are considered based on the practical hardware.
\end{itemize}
\begin{itemize}
\item[$\bullet$]
Since the formulated problem is NP hard, it is difficulty to directly solve. We first design the analog beamforming based on the RISs' locations and phase compensation principle. Next, we still need to jointly solve the digital beamforming of the BS and reflection coefficients of the RISs. We propose an alternatively optimization algorithm to deal with it. Specifically, the reflection coefficients are fixed, and we obtain the digital beamforming based on the minimum mean square error (MMSE) technique. After that, the reflection coefficients are solved via the coordinate update approach. The finally solutions are obtained via repeating the above procedure until convergence. Furthermore, the robustness of the proposed joint wideband beamforming against the impacts of imperfect CSI is analyzed.
\end{itemize}

Notations: Lower-case and upper-case boldface letters represent vectors and matrices, respectively. $(\cdot)^{T}$ and $(\cdot)^{H}$ denote the transpose and Hermitian transpose, respectively. $|\cdot|$ denotes the absolute operator. $\left\|\cdot \right\|$ is the Frobenius norm. $\lfloor \cdot \rfloor$ and $\lceil \cdot \rceil$ are the floor and ceil function, respectively. $\rm diag(\cdot)$ represents diagonal operation. $\mathbb{C}^{x \times y}$ denotes the space of $x \times y$ complex matrix. $\Re\{\cdot\}$ denotes the real part of a complex number. $\mathcal{C} \mathcal{N}\left(A, B \right)$ represents the Gaussian distribution with mean $A$ and covariance $B$.
\section{Basic System Model}
We consider a distributed RISs-aided THz communication system as shown in Fig. 1. The THz signals are poor diffraction and vulnerable to the obstruction due to its ultra-high frequency~\cite{refE1}~\cite{refE2}. Thus, the direct links between the BS and users are assumed to be blocked by buildings. We set that the BS is consisted of an $N$-antenna uniform linear array (ULA) and  $N_{{\rm{RF}}}$ $(N \geq N_{{\rm{RF}}})$ RF chains to serve $K$ single-antenna users. Let $\mathcal{R}=\{1, \cdots, R\}$ denote the index set of RISs, we assume that all RISs own the same size with $N_{\rm {RIS}}={M_{x}}\times{M_{y}}$ elements, where $M_{x}$ and $M_{y}$ represent the number of rows and columns, respectively. Let $\mathcal{M}_{x}=\{1, \cdots, M_{x}\}$ and $\mathcal{M}_{y}=\{1, \cdots, M_{y}\}$ denote the index sets of elements in rows and columns. The orthogonal frequency division multiplexing technique with total $M$ subcarriers is applied to realize the reliable wideband transmission. The frequency of the $m$-th carrier can be expressed as $f_{m}=f_{c}+\frac{B}{M}\left(m-1-\frac{M-1}{2}\right), m=1,2,\cdots, M$, where $f_c$ and $B$ are the central frequency and bandwidth, respectively.
\begin{figure}[tbp]
	\centering
		\label{complexity} 
		\includegraphics[width=9cm,height=5cm]{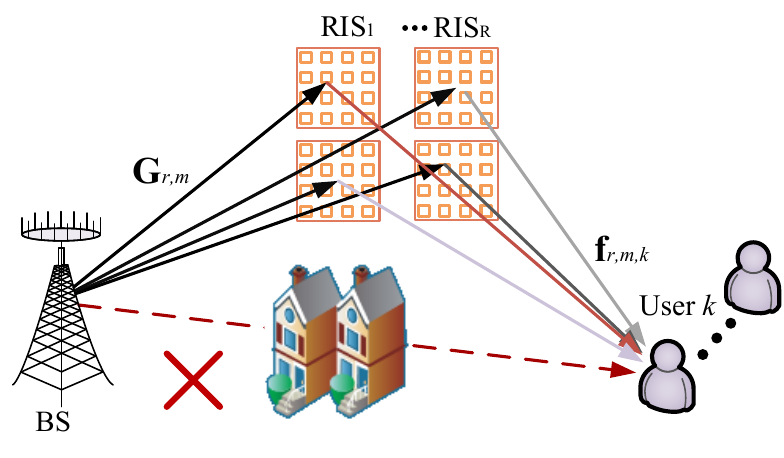}
	\caption{The distributed RISs-aided THz system.}
\end{figure}

Thus, the equivalent channel $\mathbf{h}_{m, k}$ between the BS and the $k$-th user on the $m$-th subcarrier can be expressed as
\begin{eqnarray}
  \mathbf{h}_{m, k}=\sum_{r=1}^{R} \mathbf{f}_{r, m, k} \mathbf{\Phi}_{r} \mathbf{G}_{r, m},
\end{eqnarray}
where $\mathbf{f}_{r, m, k} \in \mathbb{C}^{1 \times N_{\rm{RIS}}}$ denotes the channel vector between the $r$-th RIS and the $k$-th user on the $m$-th subcarrier, $\mathbf{G}_{r, m} \in \mathbb{C}^{N_{\rm{RIS}} \times N}$ represents the channel matrix from the BS to the $r$-th RIS on the $m$-th subcarrier. $\mathbf{\Phi}_{r}=\operatorname{diag}\left(\varphi_{r,1,1}, \cdots, \varphi_{r,m_{x},m_{y}}, \cdots, \varphi_{r,M_{x},M_{y}}\right), r \in \mathcal{R}$, $m_{x} \in \mathcal{M}_{x}, m_{y} \in \mathcal{M}_{y}$, is the diagonal reflection coefficients matrix of the $r$-th RIS with $\varphi_{r,m_{x},m_{y}}= \varepsilon_{r,m_{x},m_{y}} e^{j \phi_{r,m_{x},m_{y}}}$. To maximize the reflection efficiency, we set $\varepsilon_{r,m_{x},m_{y}}=1$ for $r \in \mathcal{R}$, $m_{x} \in \mathcal{M}_{x}, m_{y} \in \mathcal{M}_{y}$. We apply the Saleh-Valenzuela channel model~\cite{refG1}, and thus the channel matrix $\mathbf{G}_{r, m}$ can be expressed as
\begin{eqnarray}
\mathbf{G}_{r, m}=\sum_{l_{1}=1}^{L_{1}} \alpha_{l_{1}}^{r} e^{-j 2 \pi \tau_{l_{1}}^{r} f_{m}} \mathbf{b}\left(u_{l_{1}}^{r}, v_{l_{1}}^{r}\right) \mathbf{a}\left(\theta_{l_{1}}^{r}\right)^{ H},
\end{eqnarray}
where $L_{1}$ represents the number of paths, $\alpha_{l_{1}}^{r}$ and $\tau_{l_{1}}^{r}$ respectively denote the gain and delay of the $l_{1}$-th path at the $r$-th RIS. $\mathbf{a}\left(\theta_{l_{1}}^{r}\right)$ and $\mathbf{b}\left(u_{l_{1}}^{r}, v_{l_{1}}^{r}\right)$ respectively denote the array response vectors at the BS and RIS, which can be denoted as
\begin{eqnarray}
\mathbf{a}\left(\theta_{l_{1}}^{r}\right)=
\frac{1}{\sqrt{N}}\left[1, \ldots, e^{j 2 \pi d \frac{f_{m}}{c} n \sin \theta_{l_{1}}^{r}}, \ldots, e^{\left.j 2 \pi d \frac{f_{m}}{c}\left(N-1\right) \sin \theta_{l_{1}}^{r}\right)}\right]^{T},
\end{eqnarray}
and
\begin{eqnarray}
\begin{split}
\mathbf{b}\left(u_{l_{1}}^{r}, v_{l_{1}}^{r}\right)
&=\frac{1}{\sqrt{N_{\rm{RIS}}}}[1, \ldots, e^{j 2 \pi d \frac{f_{m}}{c}(m_{x} \cos u_{l_{1}}^{r} \sin v_{l_{1}}^{r}+m_{y} \cos v_{l_{1}}^{r})}, \\
&\ldots, e^{j 2 \pi d \frac{f_{m}}{c}((M_{x}-1) \cos u_{l_{1}}^{r} \sin v_{l_{1}}^{r}+(M_{y}-1) \cos v_{l_{1}}^{r})}]^{T},
\end{split}
\end{eqnarray}
where $c$ and $d$ are the speed of light and the distance between two consecutive antennas, respectively. We set $d=\lambda_{c} / 2$, and $\lambda_{c}$ is the wavelength of the central frequency $f_{c}$.  $\theta_{l_{1}}^{r} \in[-\pi / 2, \pi / 2]$ is the physical direction of the $l_{1}$-th path departing from the BS to the $r$-th RIS. $u_{l_{1}}^{r}$ and $v_{l_{1}}^{r} \in[-\pi / 2, \pi / 2]$ represent the azimuth and elevation angles of arrivals (AoAs) of the $l_{1}$-th path at the $r$-th RIS, respectively.

Next, the channel vector from the $r$-th RIS to the $k$-th user on the $m$-th subcarrier is denoted as
\begin{eqnarray}
\mathbf{f}_{r, m, k}=\sum_{l_{2}=1}^{L_{2}} \alpha_{l_{2}}^{r, k} e^{-j 2 \pi \tau_{l_{2}}^{r, k} f_{m}} \mathbf{b}\left(u_{l_{2}}^{r, k}, v_{l_{2}}^{r, k}\right),
\end{eqnarray}
where $L_{2}$ represents the number of paths, $\alpha_{l_{2}}^{r, k}$ and $\tau_{l_{2}}^{r, k}$ respectively denote the gain and delay of the $l_{2}$-th path from the $r$-th RIS to the $k$-th user, $\mathbf{b}\left(u_{l_{2}}^{r,k}, v_{l_{2}}^{r,k}\right)$ denotes the transmit array response vector at the RIS, namely
\begin{eqnarray}
\begin{split}
\mathbf{b}\left(u_{l_{2}}^{r,k}, v_{l_{2}}^{r,k}\right)=&
\frac{1}{\sqrt{N_{\rm{RIS}}}}[1, \ldots, e^{j 2 \pi d \frac{f_{m}}{c}(m_{x} \cos u_{l_{2}}^{r,k} \sin v_{l_{2}}^{r, k}+m_{y} \cos v_{l_{2}}^{r,k})},\\
&\ldots, e^{j 2 \pi d \frac{f_{m}}{c}((M_{x}-1) \cos u_{l_{2}}^{r,k} \sin v_{l_{2}}^{r, k}+(M_{y}-1) \cos v_{l_{2}}^{r,k})}]^{T},
\end{split}
\end{eqnarray}
where $u_{l_{2}}^{r,k}$ and $v_{l_{2}}^{r,k} \in[-\pi / 2, \pi / 2]$ represent the azimuth and elevation angles of departures (AoDs) of the $l_{2}$-th path from the $r$-th RIS to the $k$-th user, respectively.

The above is the basic system model for the distributed RISs-aided THz communications. Next, we first study how to mitigate the beam split effect by designing more practical antenna structure at the BS, and then investigate the joint beamforming optimization problem.
\section{Double-layer TTD Scheme and Performance Analysis}
To overcome the beam split effect, existing works mainly consider the single-layer TTD scheme. However, it requires each time delay line to provide a large-range delay, especially for a large antenna array, and TTD with large-range delay usually has high power consumption,
insertion loss, and hardware complexity, which is impractical for TTD circuits~\cite{refG21}. Therefore, to mitigate the beam split effect with low power consumption and  hardware cost, we propose a double-layer TTD scheme at the~BS.

\begin{figure*}[htbp]
\centering
    \label{RIS_subcarrier} 
    \includegraphics[width=15cm,height=6cm]{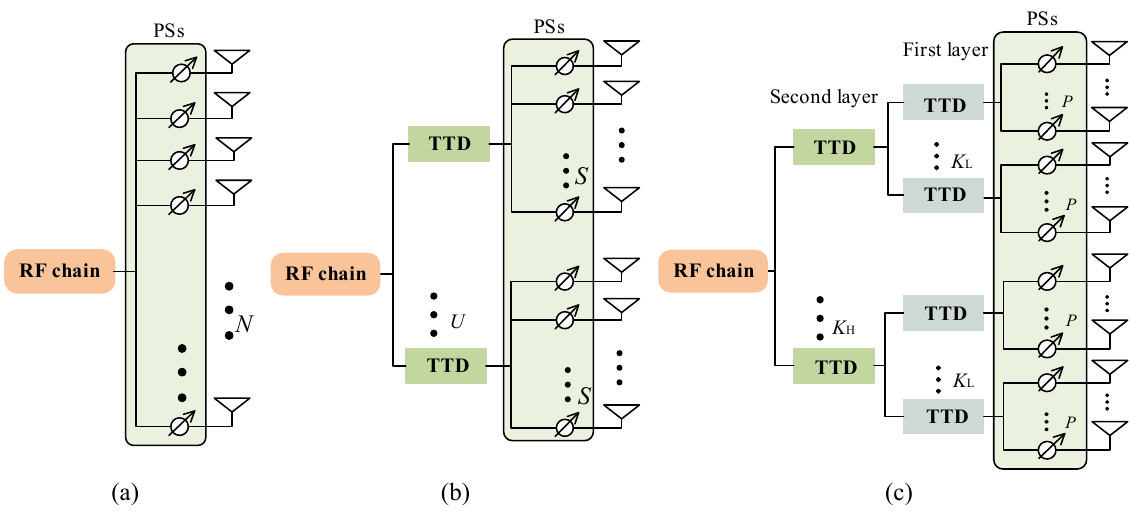}
	
	\caption{Three different antenna schemes: (a) PS scheme, (b) Single-layer TTD scheme, (c) Double-layer TTD scheme.}
\end{figure*}
\subsection{Phase Compensation Principle}
We first introduce the phase compensation principle. For convenience, we assume that there is only one RF chain connecting to $N$ antennas via PSs at the BS as shown in Fig. 2(a), and one single-antenna user is served. Although there are a few scattering components in THz communications, their power are much lower than that of the LoS component~\cite{refG3}. Therefore, we only consider the LoS component here, and thus the channel vector $\mathbf{\bar h}_{m} \in \mathbb{C}^{1 \times N}$ of the BS-user link on the $m$-th carrier can be expressed~as
\begin{eqnarray}
\mathbf{\bar h}_{m}= \alpha e^{-j 2 \pi \mathcal{\tau} f_{m}} \mathbf{a}\left(\theta_{0}\right)^{H},
\end{eqnarray}
where $\alpha$ and $\tau$ respectively denote the gain and delay, $\theta_{0} \in[-\pi / 2, \pi / 2]$ is the AoD,
$\mathbf{a}\left(\theta_{0}\right)$ denotes the array response vector at the BS, namely
\begin{eqnarray}
\mathbf{a}\left(\theta_{0}\right)=\frac{1}{\sqrt{N}}\left[1, \ldots, e^{j 2 \pi d \frac{f_{m}}{c} n \sin \theta_{0}}, \ldots, e^{\left.j 2 \pi d \frac{f_{m}}{c}\left(N-1\right) \sin \theta_{0}\right)}\right]^{T}.
\end{eqnarray}

To form a beam at direction $\theta_{0}$, the phase $\Psi_{c}$ excited by the $n$-th PS at center frequency $f_{c}$ should be~\cite{refH}
\begin{eqnarray}
  \Psi_{c}=\frac{2 \pi f_{c}}{c} (n-1) d \sin \theta_{0}=\omega_{c}\tau_{n} ,  n=1,2, \ldots, N,
\end{eqnarray}
where
\begin{eqnarray}
\tau_{n}=\frac{(n-1)d}{c} \sin \theta_{0}=(n-1)T_d \sin \theta_{0},
\end{eqnarray}
is the propagation delay between the first and $n$-th antenna, $\omega_{c}$ is the angular frequency corresponding to the center frequency $f_{c}$, and $T_d=\frac{d}{c}$ is the delay between two consecutive antennas. Thus, the analog beam  $\mathbf{f}_{\rm ps}$ under the PS scheme can be written as
\begin{eqnarray}
\mathbf{f}_{\rm ps}=\frac{1}{\sqrt{N}}\left[1, \ldots, e^{j 2 \pi f_{c} n T_d\sin \theta_{0}}, \ldots, e^{\left.j 2 \pi f_{c}\left(N-1\right) T_d\sin \theta_{0}\right)}\right]^{T}.
\end{eqnarray}
Wrapping the phase shift to $[0, 2\pi]$, we obtain
\begin{eqnarray}
  \Psi_{c}'=\Psi_{c}-\rm{TRUNC}(\frac{\Psi_{c}}{2\pi}),
\end{eqnarray}
where $\rm{TRUNC}$ is a function that represents the integral part of its argument. Due to the frequency-independent property of PSs, the phase adjusted by each PS is common for all frequencies. But when the frequency varies from $f_{c}$ to $f_{m}$, the ideal phase should be
\begin{eqnarray}
  \Psi_{p}=2 \pi f_{m}(n-1) T_d \sin \theta_{0}.
\end{eqnarray}
Thus, there is a phase difference between the ideal phase and practical phase, which can be expressed as
\begin{eqnarray}
 \Delta \Psi=2 \pi (f_{m}-f_c) (n-1) T_d \sin \theta_{0}.
\end{eqnarray}
The phase difference at $f_{m}$ leads to the beam direction moving to $\theta_{0}^{'}$, namely
\begin{eqnarray}
\theta_{0}^{'}=\text{arcsin}(\frac{f_c}{f_m}\sin \theta_{0}).
\end{eqnarray}
It is observed that the phase difference increases with the bandwidth and the number of antennas. Since the beam split essentially results from the propagation delay across the antenna array aperture, a reasonable approach to tackle this issue is to compensate the propagation delay. Consequently, the TTD is introduced to thoroughly eliminate the beam split. Fig. 2 (b) shows a typical TTD antenna structure, where $N$ antennas are uniformly divided into $U$ subarrays and each one includes $S=N/U$ antennas. Meanwhile, these antennas are connected to one TTD via $S$ PSs at each subarray, and then all TTDs are connected to the RF chain. Next, as shown in Fig. 3, we define $\tau_u$ as the signal transmission time delay difference from the first antenna of the first subarray to the first antenna of the $u$-th subarray, namely
\begin{eqnarray}
 \tau_{u}= (u-1) S T_d \sin \theta_{0},
\end{eqnarray}
where $u=1,2, \ldots U$. From (16), we have $\tau_{u} \in [0,(U-1) S T_d]$ with $\theta_{0} \in [-\pi/2, \pi/2]$. For the single-layer TTD scheme, the transmission signal phase at each antenna is jointly controlled by the frequency-dependent TTD and frequency-independent PS~\cite{refJ}. Therefore, the transmission signal phase $\Psi_{u,s}$ of the $s$-th element in the $u$-th subarray at the frequency $f_m$ can be written as
\begin{eqnarray}
\Psi_{u,s}=2 \pi f_m \left(u-1\right) S T_d \sin \theta_0+2 \pi f_c(s-1) T_d \sin \theta_0,
\end{eqnarray}
where $s=1,2, \ldots S$. The corresponding analog beamforming $\mathbf{f}_{\rm ttd}$ can be expressed as
\begin{eqnarray}
\mathbf{f}_{\rm ttd}=\frac{1}{\sqrt{N}}[1, \ldots, e^{j\Psi_{u,s}}, \ldots, e^{\Psi_{U,S}}]^{T}.
\end{eqnarray}
To achieve unbiased beam synthesis, the ideal phase at frequency $f_m$ should be
\begin{eqnarray}
\hat{\Psi}_{u,s}=2 \pi f_m[\left(u-1\right) S +(s-1)] T_d \sin \theta_0.
\end{eqnarray}
\begin{figure}[tbp]
\centering
    \label{RIS_subcarrier} 
    \includegraphics[width=9cm,height=6.5cm]{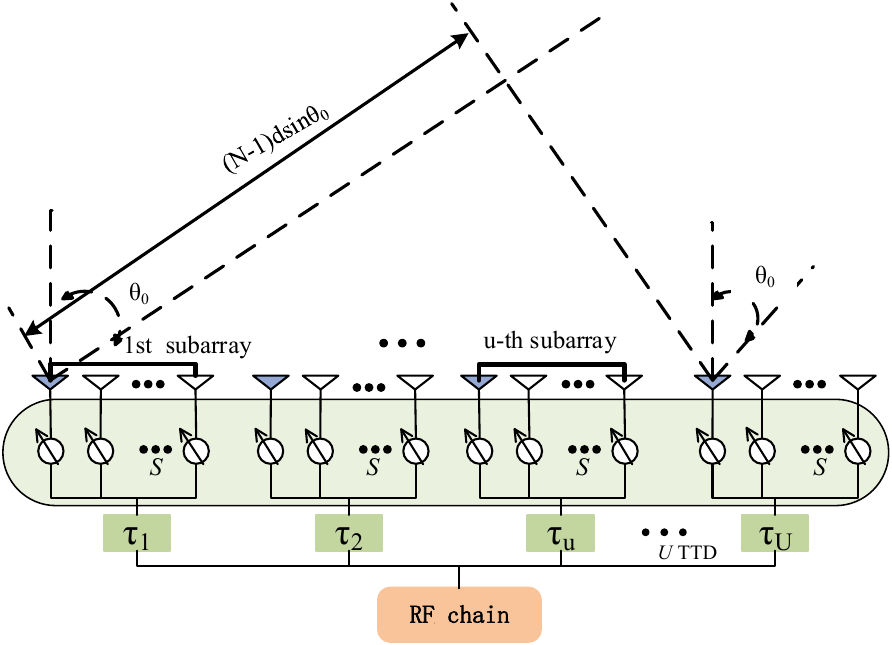}
	
	\caption{Configuration of the typical single-layer TTD scheme.}
\end{figure}

Thus, there exists a phase error under the single-layer TTD scheme, namely
\begin{eqnarray}
 \Delta \Psi_{u,s}=2 \pi (f_m-f_c)(s-1) T_d \sin \theta_0.
\end{eqnarray}
One can observe that the phase error is related to the subarray aperture and it increases with the number of elements  in each subarray. Therefore, for reducing the phase error and beam split effect, we can increase the subarray number (i.e., TTDs number), but this will cause a high hardware cost.

Next, to clearly understand the hardware implementation of TTD, we present an 8-bit TTD with a minimum delay of 1 ps~\cite{refJ} as an example. In order to construct a complete 8-bit TTD, 8 TTD elements should be cascaded for a total delay of 255 ps, with the least significant  bit (LSB) equal to 1 ps and the most significant bit (MSB) equal to 128 ps, as shown in the circuit block diagram of Fig. 4. Each discretely unit is realized by cascaded time delay units, reference units and input and output single-pole-double-throw (SPDT) switches~\cite{refC1}~\cite{refJ00},  which provide different levels of time delay and time delay  selection, respectively. Due to the cascaded structure, the  power consumption, insertion loss, and hardware complexity  of the TTD are summation of those of the time delay units, and switches. In addition, the insertion loss of the time delay is increased as the frequency increases~\cite{refJ000}. Thus, in the THz band, fewer bits can reduce the number of time delay units and switches, which further decreases the power consumption, insertion loss, and hardware complexity.
\begin{figure}[htbp]
	\centering
		\label{complexity} 
		\includegraphics[width=9.5cm,height=3.2cm]{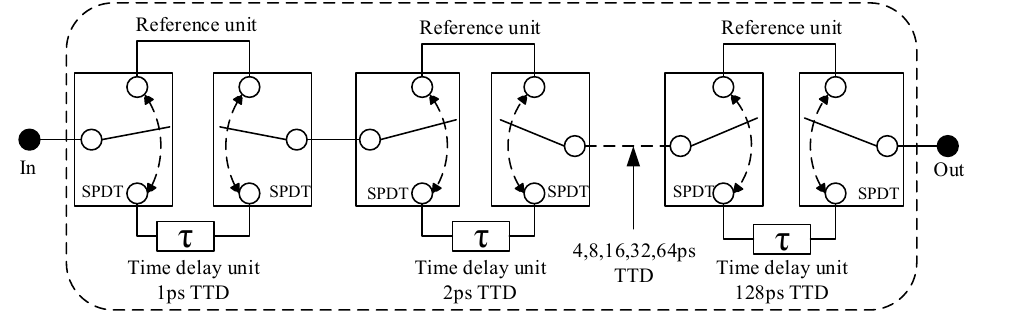}
	\caption{Circuit block diagram of the 8-bit TTD.}
\end{figure}
\subsection{Proposed Double-layer TTD Scheme}
Fig. 2 (c) is our proposed double-layer TTD scheme, where the second layer includes $K_{\rm H}$ TTDs and each TTD is connected to $K_{\rm L}$ TTDs of the first layer. Meanwhile, each TTD of the first layer is connected to $P$ antennas via PSs. We define $\tau_{k_h, k_l}$ as the signal transmission time delay difference from the first antenna of the first subarray to the first antenna of the $k_l$-th subarray related to the $k_h$-th TTD of the second layer, namely
\begin{eqnarray}
\tau_{k_h, k_l}=\left(k_l-1\right) P T_d\sin \theta_0,
\end{eqnarray}
where $k_l=1,2, \ldots K_{\rm L}$, $k_h=1,2, \ldots K_{\rm H}$. Next, we define $\tau_{k_h}$ as the signal transmission time delay difference from the first antenna connecting to the first TTD of the second layer to the first antenna connecting to the $k_h$-th TTD of the second layer, namely
\begin{eqnarray}
\tau_{k_h}=\left(k_h-1\right) K_{\rm L} P T_d \sin \theta_0.
\end{eqnarray}
From (21) and (22), we have $\tau_{k_h, k_l} \in [0,(K_{\rm L}-1) P T_d]$ and $\tau_{k_h} \in [0,(K_{\rm H}-1) K_{\rm L} P T_d]$ with $\theta_{0} \in [-\pi/2, \pi/2]$, $1\leq k_l\leq K_L$ and $1\leq k_h\leq K_{\rm H}$. Therefore, for the double-layer TTD scheme, the time delay range of the TTD at the first layer is always smaller than that of the TTD at the second layer. And the phase of each antenna is controlled by the double-layer network and PSs. The corresponding phase value $\Psi_{h,l,p}$ of the $p$-th element in the $k_l$-th TTD under the $k_h$-th subarray at the frequency $f_m$ is
\begin{eqnarray}
\begin{split}
\Psi_{h,l,p}&=2 \pi f_m \left(k_h-1\right) K_{\rm L} P T_d\sin \theta_0+\\
&2 \pi f_m\left(k_l-1\right) P T_d \sin \theta_0+2 \pi f_c(p-1) T_d \sin \theta_0,
\end{split}
\end{eqnarray}
where $p=1,2, \ldots P$.
The phase error of each element compared to the ideal phase is
\begin{eqnarray}
 \Delta \Psi_{h,l,p}=2 \pi (f_m-f_c) (p-1) T_d \sin \theta_0.
\end{eqnarray}
The corresponding analog beamforming $\mathbf{f}_{\rm mttd}$
excited by the double-layer TTD scheme can be expressed as
\begin{eqnarray}
\mathbf{f}_{\rm mttd}=\frac{1}{\sqrt{N}}[1, \ldots, e^{j\Psi_{h,l,p}}, \ldots, e^{\Psi_{H,L,P}}]^{T}.
\end{eqnarray}
Next, we respectively derive the normalized array gains under three different schemes. For the target direction $\theta_{0}$ on the frequency $f_m$, the array gain under the traditional PS scheme can be calculated as
\begin{align}
g_{\rm ps} \left(f_m, \theta_0\right)&=\left|\mathbf{a}^{\rm H} \mathbf{f}_{\rm ps}\right|\nonumber\\
&=\frac{1}{N}\left|\sum_{n=1}^{N} e^{-j 2 \pi d \frac{f_m}{c} (n-1) \sin \theta_0} e^{j 2 \pi d \frac{f_c}{c} (n-1) \sin \theta_0}\right|\nonumber\\
&=\frac{1}{N}\left|\Xi_{N}\left(\left(\zeta_m-1\right) \sin \theta_0\right)\right|,
\end{align}
where $\zeta_m=\frac{f_m}{f_c}$ denotes the relative frequency, and $\Xi_{N}\left(x \right)=\frac{\sin \left(\frac{\pi N}{2}x \right)}{\sin \left(\frac{\pi}{2}x \right)}$ is the Dirichlet
Sinc function~\cite{refJ0}~\cite{refJ1}. Next, the array gain under the single-layer TTD scheme should be
\begin{eqnarray}
\begin{aligned}
g_{\rm ttd} \left(f_m, \theta_0\right)&=\left|\mathbf{a}^{\rm H} \mathbf{f}_{\rm ttd}\right|\\
&=\frac{1}{N}\left|\sum_{u=1}^{U}\sum_{s=1}^{S}  e^{j 2 \pi d \frac{f_c}{c} (s-1) \sin \theta_0} e^{j \frac{2 \pi f_m}{c}\left(u-1\right) S d \sin \theta_0} \right.\\ & \left.
e^{-j 2 \pi d \frac{f_{m}}{c}[(u-1)S+(s-1)] \sin \theta_{0}}\right|\\
&=\frac{U}{N}\left|\Xi_{S}\left(\left(\zeta_m-1\right) \sin \theta_0\right)\right|.
\end{aligned}
\end{eqnarray}
Finally, the array gain under the double-layer TTD scheme can be expressed as
\begin{align}
\!\!\!\!\!\!g_{\rm mttd} \left(f_m, \theta_0\right)&=\left|\mathbf{a}^{\rm H} \mathbf{f}_{\rm mttd}\right|\nonumber\\
&=\frac{1}{N}\left|\sum_{k_h=1}^{K_{\rm H}}\sum_{k_l=1}^{K_{\rm L}}\sum_{p=1}^{P} e^{j \pi\frac{f_m}{f_c}(k_h-1)K_{\rm L} P \sin \theta_0} e^{j \pi\frac{f_m}{f_c}(k_l-1) P \sin \theta_0}\nonumber \right.\\ & \left.
e^{j \pi(p-1) \sin \theta_0} e^{-j\pi \frac{f_{m}}{f_c}[(k_h-1)K_{\rm L} P+(k_l-1)P+(p-1)] \sin \theta_{0}}\right|\nonumber\\
&=\frac{K_{\rm H} K_{\rm L}}{N}\left|\sum_{p=1}^{P}  e^{-j \pi(p-1) (\frac{f_m}{f_c}-1)\sin \theta_0}\right|\nonumber\\
&=\frac{K_{\rm H} K_{\rm L}}{N}\left|\Xi_{P}\left(\left(\zeta_m-1\right) \sin \theta_0\right)\right|.
\end{align}

From (26)-(28), one can observe that the difference of the array gains under three different schemes mainly own to their different elements number of each subarray. In the proposed double-layer TTD scheme, by introducing an additional small-delay TTD network, the required number of large-range delay TTDs can be effectively reduced. Fig. 5 shows the phase compensation of each antenna under different schemes, and we set $f_c=300$ GHz, $B=30$ GHz, $\theta_0=\pi/4$ and $N=128$.
\begin{figure}[htbp]
	\centering
		\label{complexity} 
		\includegraphics[width=9.5cm,height=6.5cm]{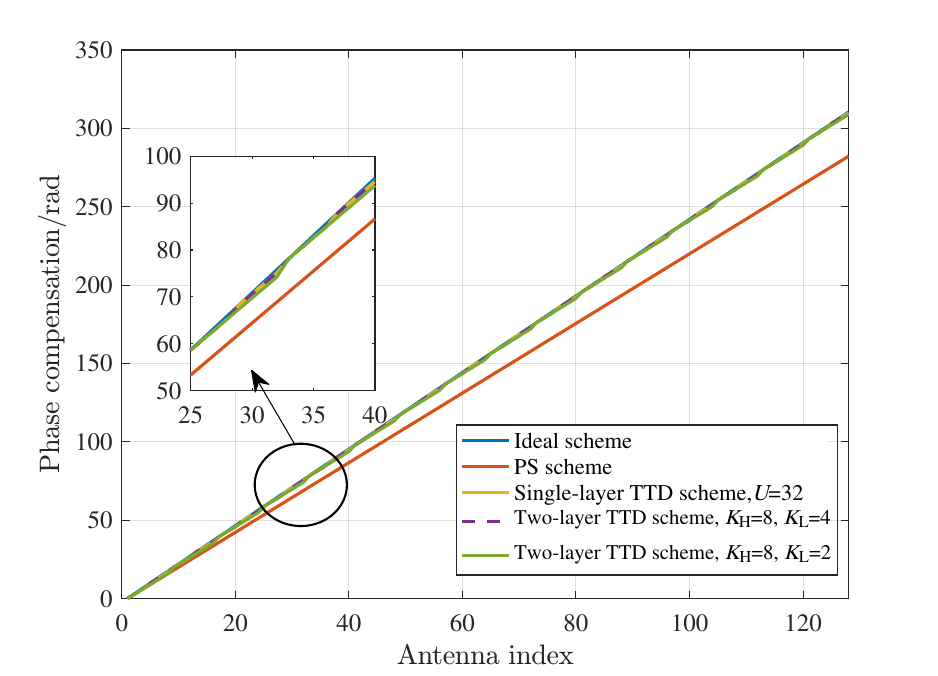}
	\caption{Phase compensation of each antenna.}
\end{figure}
\begin{figure}[htbp]
	\centering
		\label{complexity} 
		\includegraphics[width=9.5cm,height=6.5cm]{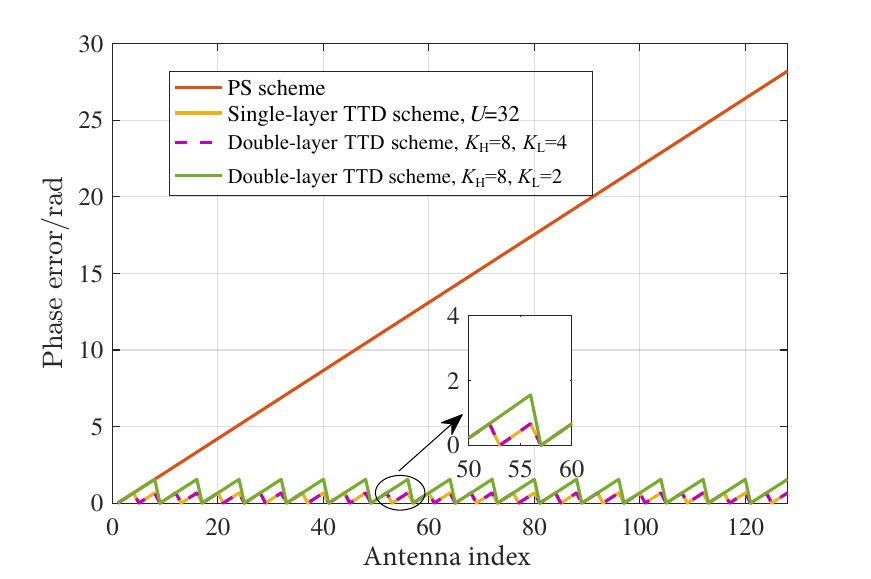}
	\caption{ Phase error of each antenna.}
\end{figure}
In the single-layer TTD scheme, we set the number of TTDs $U=32$. In the double-layer TTD scheme, we consider two configurations, i.e., $K_{\rm H}=8, K_{\rm L}=4$ and $K_{\rm H}=8, K_{\rm L}=2$. Without considering the quantization error of the TTD device, we can find that the phase compensation under the single- and double-layer TTD schemes is almost the same for $K_{\rm H}=8, K_{\rm L}=4$. However, the required number of large-range delay TTDs in the double-layer TTD scheme is sharply reduced. When $K_{\rm H}=8, K_{\rm L}=2$, the performance is slightly lower, but the total number of TTDs is further reduced.
Fig. 6 depicts phase error of each antenna under different schemes. One can observe that the phase error under the single- and double-layer TTD schemes is periodic and the period becomes shorter when the number of subarray antennas is smaller. Nevertheless, the phase error under the PS scheme increases linearly antenna index.
\begin{figure}[htbp]
	\centering
		\label{complexity} 
		\includegraphics[width=9.5cm,height=6.5cm]{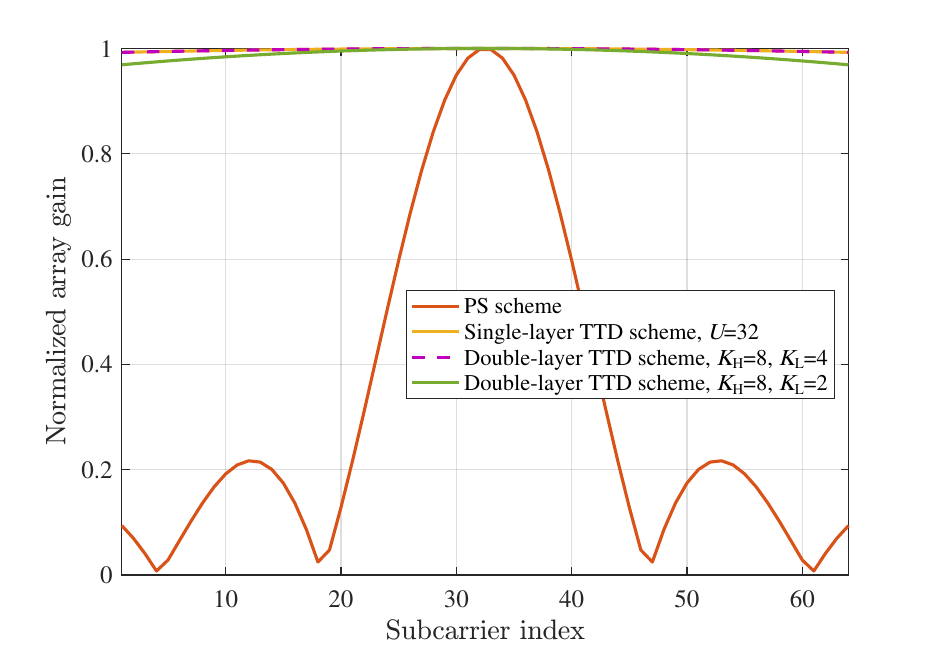}
	\caption{Normalized array gain.}
\end{figure}

Fig. 7 illustrates the normalized array gain of each subcarrier under different schemes. It can be observed that the gain under single- and double-layer TTD schemes can almost obtain the high performance across the entire bandwidth. Although the gain under the double-layer TTD scheme with $K_{\rm H}=8$ and $K_{\rm L}=2$ is a little lower, the number of large-range delay TTDs is smaller. Whereas, the gain loss under the PS scheme is largest, which seriously effects the system performance.

Furthermore, considering the practical hardware limitation, the TTD can only realize the limited discrete time delays. Therefore, we assume that the TTD under the first layer and second layer owns $2^{P_{\rm L}}$ and $2^{P_{\rm H}}$ discrete values, respectively, where $P_{\rm L}$ and $P_{\rm H}$ denote the bit number. Thus, the set of discrete time delay $\tau_{k_{h},k_{l}}$ and $\tau_{k_{h}}$ can be given by
\begin{eqnarray}
\tau_{k_h, k_{l}} \in \mathcal{T}_{1}=\{0, D, 2D, \cdots, (2^{P_{\rm L}}-1)D\},
\end{eqnarray}
\begin{eqnarray}
\tau_{k_{h}} \in \mathcal{T}_{2}=\{0, D, 2D, \cdots, (2^{P_{\rm H}}-1)D\},
\end{eqnarray}
where $D$ is the time delay step.

Without loss of generality, we first consider the time delay step as $D=T_{c}$. To ensure that the maximum time delay interval between $P$ array elements is within $T_c$ and each array element can obtain the required optimal time delay, $P$, $P_{\rm L}$ and $P_{\rm H}$ should satisfy the following conditions~\cite{refJ11}
\begin{subequations}\label{OptA}
\begin{align}
&\tau_{n+P-1}\left(\theta_0\right)-\tau_n\left(\theta_0\right)<T_c \label{OptA0},\\
&\tau_{n+P\left(K_{\rm L}-1\right)}\left(\theta_0\right)-\tau_n\left(\theta_0\right)<\left(2^{P_{\rm L}}-1\right) T_c \label{OptA1},\\
&\tau_{N-P K_{\rm L}}\left(\theta_0\right)-\tau_0\left(\theta_0\right)<\left(2^{P_{\rm H}}-1\right) T_c.
\end{align}
\end{subequations}
Then, the required bit at each layer and minimum array elements are calculated as
\begin{subequations}\label{OptA}
\begin{align}
&P \leq \lfloor \frac{T_c}{T_d \sin \theta_0}\rfloor\label{OptA0},\\
&P_{\rm L} \geq \lceil \log _2 \frac{\left(K_{\rm L}-1\right)P T_d \sin \theta_0}{T_c}\rceil\label{OptA1},\\
&P_{\rm H} \geq \lceil\log _2 \frac{\left(N-P K_{\rm L}\right) T_d \sin \theta_0}{T_c}\rceil,
\end{align}
\end{subequations}
where $\lfloor x \rfloor$ and $\lceil x \rceil$ are the floor and ceil function, respectively. If $P_{\rm L}=0$ and $K_{\rm L}=1$, double-layer TTD scheme degrades to the single-layer TTD scheme. The required bit $P_s$ for the single-layer TTD  should be
\begin{eqnarray}
P_s \geq \lceil\log _2 \frac{\left(U-1\right)S T_d \sin \theta_0}{T_c}\rceil.
\end{eqnarray}
Next, we compare these two schemes. It is obvious that the hardware complexity of the TTD device and the antenna system mainly depends on the number of bits in an individual TTD device and the total number of TTDs, respectively. Consequently, the bit ratio can be used to measure the degree of reduction in hardware cost of the double-layer TTD scheme relative to the single-layer TTD scheme, which is defined as
\begin{eqnarray}
\eta=\frac{(K_{\rm H} P_{\rm H} + K_{\rm H} K_{\rm L} P_{\rm L})}{U P_s}.
\end{eqnarray}
We assume $N=128$, $f_c=300$ GHz, $\theta_0= \pi/4$, $U=32$, $K_{\rm H}=8$, $K_{\rm L}=4$, and then have $P_s \geq 6$, $P_{\rm H} \geq 6$, $P_{\rm L} \geq 3$ according to (32) and (33). To minimize the hardware cost, we set $P_s = 6$, $P_{\rm H} = 6$ and $P_{\rm L} = 3$. Then, we can calculate that the total bits under single- and double-layer TTD schemes are $B_s=U P_s=192$ and $B_{m}=K_{\rm H} P_{\rm H} + K_{\rm H} K_{\rm L} P_{\rm L}=144$, respectively. Therefore, the bit ratio is $\eta=75\%$. By introducing an additional small-range delay TTD network, it is obvious that the total required bits of TTD and the number of large-range delay TTDs are reduced under the double-layer TTD scheme, and thus hardware cost is reduced effectively.
\section{Problem Formulation and Solutions}
In this section, we investigate the joint beamforming optimization problem for the distributed RISs-aided THz communications based on the proposed double-layer TTD scheme. Based on the practical hardware limitation, we consider the finite-resolution phase shift and time delay. We assume that the number of RF chains $N_{\rm{RF}}$ is equal to the number of RISs, i.e., $N_{\rm{RF}}=R$, and the fully connect structure is considered as shown in Fig. 8.
\begin{figure}[t]
	\centering
		\label{complexity} 
		\includegraphics[width=9.5cm,height=7.5cm]{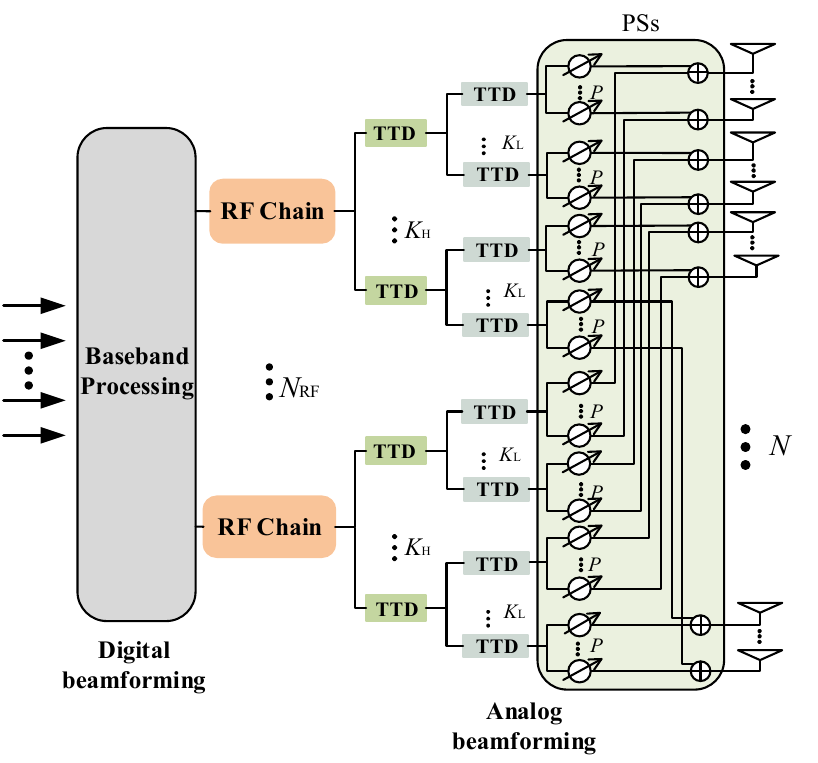}
	\caption{Fully connect antenna structure with double-layer TTD scheme.}
\end{figure}

\subsection{Problem Formulation}
The received signal of the $k$-th user on the $m$-th subcarrier can be written as
\begin{eqnarray}
y_{m, k}=
\mathbf{h}_{m, k} \mathbf{F} \mathbf{d}_{m, k} {s}_{m, k}+\sum_{j=1, j \neq k}^{K} \mathbf{h}_{m,k} \mathbf{F} \mathbf{d}_{m, j} s_{m, j}+n_{m, k},
\end{eqnarray}
where $\mathbf{F}=\mathbf{F}_{\rm{A}} \mathbf{F}_{\rm{L}}\mathbf{F}_{\rm{H}}$ is the analog beamforming matrix realized by double-layer TTD network and PSs.
$\mathbf{F}_{\rm{A}}\in \mathbb{C}^{N \times K_{\rm{L}}K_{\rm{H}}N_{\rm{RF}}}= \operatorname{diag}([\mathbf{F}_{1}, \cdots, \mathbf{F}_{n_{\rm rf}}, \cdots,\mathbf{F}_{N_{\rm{RF}}}])$, where $\mathbf{F}_{n_{\rm rf}}\in \mathbb{C}^{P K_{\rm{L}}K_{\rm{H}}\times K_{\rm{L}}K_{\rm{H}}}= \operatorname{diag}([\mathbf{c}_{n_{\rm rf},1,1},\cdots,\mathbf{c}_{n_{\rm rf},k_{h},k_{l}}, \cdots,\mathbf{c}_{n_{\rm rf},K_{\rm{H}}, K_{\rm{L}}}])$, and $\mathbf{c}_{n_{\rm rf},k_{h},k_{l}}\in \mathbb{C}^{P \times 1}$ denotes the beamforming vector generated by $P$ PSs connecting to the $n_{\rm rf}$-th RF chain
via the $k_{l}$-th subarray related to the $k_{h}$-th TTD of the second layer.
$\mathbf{F}_{\rm{L}}\in \mathbb{C}^{ K_{\rm{L}}K_{\rm{H}}N_{\rm{RF}}\times K_{\rm{H}}N_{\rm{RF}}}$ denotes the frequency-dependent phase shifts realized by the first-layer TTD network, and satisfies
\begin{eqnarray}
\mathbf{F}_{\rm{L}}=\operatorname{diag}\left(\left[e^{j 2 \pi f_{m} \boldsymbol{\tau}_{1,1}}, e^{j 2 \pi f_{m} \boldsymbol{\tau}_{1,2}},\cdots, e^{j 2 \pi f_{m} \boldsymbol{\tau}_{N_{\rm{RF}},K_{\rm{H}}}}\right]\right),
\end{eqnarray}
where $\boldsymbol{\tau}_{n_{\rm rf},k_{h}} \in \mathbb{C}^{K_{\rm{L}} \times 1}=\left[\tau_{n_{\rm rf},k_{h}, 1}, \tau_{n_{\rm rf}, k_{h}, 2}, \cdots, \tau_{n_{\rm rf},k_{h}, K_{\rm{L}}}\right]^{\rm T}$ is the time delay vector realized by $K_{\rm{L}}$ TTD elements connecting to the $k_h$-th TTD  of the second layer under the $n_{\rm rf}$-th RF chain.
$\mathbf{F}_{\rm{H}}\in \mathbb{C}^{K_{\rm{H}}N_{\rm{RF}}\times N_{\rm{RF}}}$ denotes the frequency-dependent phase shifts realized by the second-layer TTD network, and satisfies
\begin{eqnarray}
\mathbf{F}_{\rm{H}}=\operatorname{diag}\left(\left[e^{j 2 \pi f_{m} \boldsymbol{\tau}_{1}}, e^{j 2 \pi f_{m} \boldsymbol{\tau}_{2}},\cdots, e^{j 2 \pi f_{m} \boldsymbol{\tau}_{N_{\rm{RF}}}}\right]\right),
\end{eqnarray}
where $\boldsymbol{\tau}_{n_{\rm rf}} \in \mathbb{C}^{K_{\rm{H}} \times 1}=\left[\tau_{n_{\rm rf}, 1}, \tau_{n_{\rm rf}, 2}, \cdots, \tau_{n_{\rm rf}, K_{\rm{H}}}\right]^{\rm T}$ is the time delay vector realized by the second-layer TTD network connecting to the $n_{\rm rf}$-th RF chain.
 In addition,
$\mathbf{d}_{m, k}\in \mathbb{C}^{N_{\rm{RF}} \times 1}$ denotes the digital beamforming vector. $n_{m,k} \sim \mathcal{C} \mathcal{N}\left(0, \sigma_{m,k}^{2}\right)$  is the additive zero average white Gaussian noise (AWGN) with variance of $\sigma_{m,k}^{2}$ at the $k$-th user on the $m$-th subcarrier, and $s_{m, k}$ denotes the transmit symbol for the $k$-th user on the $m$-th subcarrier with $E\left[\left|s_{m,k}\right|^{2}\right]=1$.

Then, the SINR of the $k$-th user on the $m$-th subcarrier can be calculated as
\begin{eqnarray}
\gamma_{m,k}=\frac{\left|\mathbf{h}_{m, k} \mathbf{F} \mathbf{d}_{m, k}\right|^{2}}{\sum_{j=1, j \neq k}^{K}\left|\mathbf{h}_{m, k} \mathbf{F} \mathbf{d}_{m, j}\right|^{2}+\sigma_{m, k}^{2}},
\end{eqnarray}
and the achievable sum rate $R_{\rm{sum}}$ can be expressed by
\begin{eqnarray}
R_{\rm{sum}}=\sum_{k=1}^{K} \sum_{m=1}^{M} \log _{2}\left(1+\gamma_{m, k}\right).
\end{eqnarray}
Let $b$ denote the number of bits, and thus $2^b$ indicates the number of phase shift levels. Then, the set of discrete phase shifts generated by PSs can be expressed as
\begin{eqnarray}
\mathcal{C}=\frac{1}{\sqrt{P}}\{e^{j 0}, e^{j{\frac {2 \pi} { 2^{b}}}}, \cdots, e^{j{\frac {2 \pi} { 2^{b}}\left(2^{b}-1\right)}}\}.
\end{eqnarray}
Similarly, the set of discrete reflection coefficient of RISs can be written as
\begin{eqnarray}
\begin{split}
\mathcal {F}=\{e^{j 0}, e^{j{\frac {2 \pi} { 2^{Q}}}}, \cdots, e^{j{\frac {2 \pi} { 2^{Q}}\left(2^{Q}-1\right)}}\},
\end{split}
\end{eqnarray}
where $Q$ indicates the bit number. Finally, we formulate the joint beamforming optimization problem as follows
\begin{subequations}\label{OptA}
\begin{align}
\mathrm{P1}:&\max _{\boldsymbol{\Theta}, \mathbf{F}_{\rm{A}}, \mathbf{F}_{\rm{L}}, \mathbf{F}_{\rm{H}}, \mathbf{d}_{m, k}} R_{\rm{sum}}\label{OptA0}\\
{\rm{s.t.}}\;\;&\sum_{\mathrm{k}=1}^{K} \sum_{m=1}^{M}\left\|\mathbf{F}_{\rm{A}} \mathbf{F}_{\rm{L}}\mathbf{F}_{\rm{H}} \mathbf{d}_{m, k}\right\|^{2} \leq P_{\text {max}} \label{OptA1},\\
&\tau_{n_{\rm rf},k_{h}} \in \{0, D, 2D, \cdots, (2^{P_{\rm{H}}}-1)D\},\\
&\tau_{n_{\rm rf},k_{h},k_{l}} \in \{0, D, 2D, \cdots, (2^{P_{\rm{L}}}-1)D\},\\
&\mathbf{c}_{n_{\rm rf}, k_{h}, k_{l}}\in \frac{1}{\sqrt{P}}\{e^{j 0}, e^{j{\frac {2 \pi} { 2^{b}}}}, \cdots, e^{j{\frac {2 \pi} { 2^{b}}\left(2^{b}-1\right)}}\}\label{OptA2},\\
&\varphi_{r, m_{x},m_{y}} \in \{e^{j 0}, e^{j{\frac {2 \pi} { 2^{Q}}}}, \cdots, e^{j{\frac {2 \pi} { 2^{Q}}\left(2^{Q}-1\right)}}\},
\end{align}
\end{subequations}
where $\boldsymbol{\Theta}=\operatorname{diag}\left(\boldsymbol{\Phi}_{1}, \ldots, \boldsymbol{\Phi}_{\rm{R}}\right)$ and $P_{\text {max}}$ is the maximum available transmit power. (42b) is the total transmit power constraint, (42c) and (42d) are the discrete time delay constraint for each TTD, (42e) is the phase shift constraint for each PS, and (42f) represents the discrete reflection coefficient constraint for each RIS element. P1 aims to maximize the achievable
rate by jointly optimizing the reflection coefficient matrix $\boldsymbol{\Theta}$, frequency-independent beamforming matrix $\mathbf{F}_{\rm{A}}$, frequency-dependent phase shifts $\mathbf{F}_{\rm{H}}$ and $\mathbf{F}_{\rm{L}}$, and digital beamforming vector $\mathbf{d}_{m, k}$. Note that the constraint in (42b) is non-convex due to the coupling of $\mathbf{F}_{\rm{A}}$, $\mathbf{F}_{\rm{H}}$, $\mathbf{F}_{\rm{L}}$, and $\mathbf{d}_{m, k}$. Furthermore, the constraints in (42c)-(42f) restrict the optimization parameters to be discrete values. Thus, P1 is generally NP-hard, and there is no standard method to obtain its globally optimal solution efficiently. Next, we propose an effective algorithm to deal with~it.
\subsection{Problem Solution}
In this section, we propose a joint beamforming framework to solve P1. Firstly, based on RISs' locations, we design the analog beamforming, including phase shifts and time delays. Then, we propose an alternatively optimization algorithm to obtain the digital beamforming and reflection coefficients.
\subsubsection{Design of Analog Beamforming $\mathbf{F}$}
The analog beamforming is defined as $\mathbf{F}=\mathbf{F}_{\rm{A}} \mathbf{F}_{\rm{L}}\mathbf{F}_{\rm{H}}$, and thus we need to design $\mathbf{F}_{\rm{A}}$, $\mathbf{F}_{\rm{L}}$ and $\mathbf{F}_{\rm{H}}$ via optimizing the phase shifts and time delays. To compensate the severe array gain loss caused by the beam split, the analog beamforming should generate beams aligned with the target physical direction at all subcarriers. The key idea of the double-layer TTD scheme is that the time delays are elaborately designed to make beams over different subcarrier frequencies toward the target's physical direction, and PSs are used to compensate for the remaining phase shift of the TTD network to generate beams aligned with target's physical~direction at the central~frequency.

Therefore, we first design the time delays via the double-layer TTD network. Based on the analysis in Sec. III-B, the optimal time delay of the $k_l$-th TTD at the first layer connected to the $k_h$-th TTD at the second layer under the $n_{\rm rf}$-th RF chain can be calculated as
\begin{eqnarray}
\tau_{n_{\rm rf}, k_h, k_l}=\left(k_l-1\right) P T_d\sin \theta_{l}^{r}.
\end{eqnarray}
Similarly, the optimal time delay of the $k_h$-th TTD at the second layer connected to the $n_{\rm rf}$-th RF chain can be expressed as
\begin{eqnarray}
\tau_{n_{\rm rf}, k_h}=\left(k_h-1\right) K_{\rm L} P T_d \sin \theta_{l}^{r},
\end{eqnarray}
where $\theta_{l}^{r}$ is the physical direction of the LoS path from the BS to the $r$-th RIS. Then, we can obtain the discrete time delay $\tau_{n_{\rm rf}, k_{h}, k_{l}}^{'}$ and $\tau_{n_{\rm rf}, k_{h}}^{'}$ according to the following approximation,
\begin{eqnarray}
\tau_{n_{\rm rf}, k_{h}, k_{l}}^{'}=\underset{\tau_1 \in \mathcal{T}_{1}}{\operatorname{argmin}}\left|\tau_{n_{\rm rf}, k_{h}, k_{l}}- \tau_1\right|,
\end{eqnarray}
\begin{eqnarray}
\tau_{n_{\rm rf}, k_{h}}^{'}=\underset{\tau_2 \in \mathcal{T}_{2}}{\operatorname{argmin}}\left|\tau_{n_{\rm rf}, k_{h}}- \tau_2\right|.
\end{eqnarray}
After obtaining the time delay $\tau_{n_{\rm rf}, k_{h}, k_{l}}^{'}$, the time delay vector $\hat {\boldsymbol{\tau}}_{n_{\rm rf},k_{h}}$ realized by $K_{\rm{L}}$ TTD elements connecting to the $k_h$-th TTD of the second layer under the $n_{\rm rf}$-th RF chain is given by
\begin{eqnarray}
\hat {\boldsymbol{\tau}}_{n_{\rm rf},k_{h}} \in \mathbb{C}^{K_{\rm{L}} \times 1}=\left[\tau_{n_{\rm rf},k_{h}, 1}^{'}, \tau_{n_{\rm rf}, k_{h}, 2}^{'}, \cdots, \tau_{n_{\rm rf},k_{h}, K_{\rm{L}}}^{'}\right]^{T}.
\end{eqnarray}
The corresponding frequency-dependent phase shifts $\mathbf{F}_{\rm{L}}$ realized by the first layer is calculated as
\begin{eqnarray}
\mathbf{F}_{\rm{L}}=\operatorname{diag}\left(\left[e^{j 2 \pi f_{m} \hat {\boldsymbol{\tau}}_{1,1}}, e^{j 2 \pi f_{m} \hat {\boldsymbol{\tau}}_{1,2}},\cdots, e^{j 2 \pi f_{m} \hat {\boldsymbol{\tau}}_{N_{\rm{RF}},K_{\rm{H}}}}\right]\right).
\end{eqnarray}
Similarly, the time delay vector $\hat {\boldsymbol{\tau}}_{n_{\rm rf}}$ realized by $K_{\rm{H}}$ TTDs connecting to the $n_{\rm rf}$-th RF chain can be expressed~as
\begin{eqnarray}
\hat {\boldsymbol{\tau}}_{n_{\rm rf}} \in \mathbb{C}^{K_{\rm{H}} \times 1}=\left[\tau_{n_{\rm rf}, 1}^{'}, \tau_{n_{\rm rf}, 2}^{'}, \cdots, \tau_{n_{\rm rf}, K_{\rm{H}}}^{'}\right]^{T}.
\end{eqnarray}
The frequency-dependent phase shifts $\mathbf{F}_{\rm{H}}$ realized by the second layer is formulated as
\begin{eqnarray}
\mathbf{F}_{\rm{H}}=\operatorname{diag}\left(\left[e^{j 2 \pi f_{m} \hat {\boldsymbol{\tau}}_{1}}, e^{j 2 \pi f_{m} \hat {\boldsymbol{\tau}}_{2}},\cdots, e^{j 2 \pi f_{m} \hat {\boldsymbol{\tau}}_{N_{\rm{RF}}}}\right]\right).
\end{eqnarray}
Here, the design of the time delay only depend on the RISs's locations and BS antenna structure.

Then, we present the frequency-independent beamforming matrix $\mathbf{F}_{\rm{A}}$, which is realized by PSs. The PSs can be used to compensate for the remaining phase shift of the former TTD network and generate beams aligned with RISs' physical directions.
The beamforming vector $\mathbf{c}_{n_{\rm rf}, k_{h},k_{l}}$ generated by $P$ PSs connecting to the $n_{\rm rf}$-th RF chain via the $k_{l}$-th TTD and $k_{h}$-th TTD is given by
\begin{flalign}
\!\!\!\!\!\mathbf{c}_{n_{\rm rf}, k_{h},k_{l}}=\frac{1}{\sqrt{P}}\left[1, \ldots, e^{j 2 \pi f_{c} p T_d\sin \theta_{l}^{r}}, \ldots, e^{\left.j 2 \pi f_{c}T_d\left(P-1\right) T_d\sin \theta_{l}^{r}\right)}\right]^{T}.
\end{flalign}
Similarly, we obtain the discrete $\mathbf{c}_{n_{\rm rf}, k_{h},k_{l}}^{'}(p), p=1,2, \cdots, P$ according to the following approximation
\begin{eqnarray}
\mathbf{c}_{n_{\rm rf}, k_{h},k_{l}}^{'}(p)=\underset{c \in \mathcal{C}}{\operatorname{argmin}}\left|\mathbf{c}_{n_{\rm rf}, k_{h},k_{l}}(p)- c\right|,
\end{eqnarray}
where $c$ is element in set $\mathcal{C}$.
Then, $\mathbf{F}_{n_{\rm rf}}$ can be expressed as
\begin{eqnarray}
\mathbf{F}_{n_{\rm rf}}=\operatorname{diag}\left([\underbrace{\mathbf{c}_{n_{\rm rf}, 1,1}^{'}, \mathbf{c}_{n_{\rm rf},1,2}^{'}, \cdots, \mathbf{c}_{n_{\rm rf}, k_{\rm H},k_{\rm L}}^{'}}_{K_{\mathrm{H}}K_{\mathrm{L}} \text { columns }}]\right).
\end{eqnarray}
Finally, we can obtain the frequency-independent beamforming matrix $\mathbf{F}_{\rm{A}}$, namely
\begin{eqnarray}
\mathbf{F}_{\rm{A}}=\rm diag([\mathbf{F}_{1}, \cdots, \mathbf{F}_{n_{\rm rf}}, \cdots,\mathbf{F}_{N_{\rm{RF}}}]).
\end{eqnarray}

\subsubsection{Optimization of $\mathbf{d}_{m, k}$ with Fixed $\mathbf{\Theta}$}
So far, we obtain the analog beamforming $\mathbf{F}$, and next we solve the digital beamforming and reflection coefficients. Based on the obtained $\mathbf{F}$, the original problem P1 can be reformulated as follows
\begin{subequations}\label{OptA}
\begin{align}
\mathrm{P2}:&\max _{\boldsymbol{\Theta}, \mathbf{d}_{m, k}} R_{\rm{sum}}\label{OptA0}\\
{\rm{s.t.}}\;\;&\sum_{\mathrm{k}=1}^{K} \sum_{m=1}^{M}\left\|\mathbf{F}_{\rm{A}} \mathbf{F}_{\rm{L}}\mathbf{F}_{\rm{H}} \mathbf{d}_{m, k}\right\|^{2} \leq P_{\text {max}} \label{OptA1},\\
&\varphi_{r, m_{x},m_{y}} \in \{e^{j 0}, e^{j{\frac {2 \pi} { 2^{Q}}}}, \cdots, e^{j{\frac {2 \pi} { 2^{Q}}\left(2^{Q}-1\right)}}\}.
\end{align}
\end{subequations}
However, P2 is still difficult  to solve due to the non-convex complex objective function. Next, we propose an alternatively optimization algorithm to deal with it.

Firstly, for given the reflection coefficient matrix $\mathbf{\Theta}$, we propose an iterative algorithm based on the MMSE technique to obtain the digital beamforming $\mathbf{d}_{m, k}$. The equivalent channel vectors for the $k$-th user on the $m$-th subcarrier can be written as $\hat{\mathbf{h}}_{m, k}=\mathbf{h}_{m, k}\mathbf{F}$. Besides, based on the extension of the Sherman-Morrison-Woodbury formula~\cite{refO}, we have
\begin{eqnarray}
\left(1+\gamma_{m, k}\right)^{-1}=1-\frac{\left|\hat{\mathbf{h}}_{m, k} \mathbf{d}_{m, k}\right|^{2}}{\sum_{j=1}^{K}\left|\hat{\mathbf{h}}_{m, k} \mathbf{d}_{m, j}\right|^{2}+\sigma_{m, k}^{2}}.
\end{eqnarray}
The MMSE-receive combining filter at the $k$-th user on the $m$-th subcarrier is given as
\begin{eqnarray}
\chi_{m, k}^{*}=\rm arg \min _{\chi_{m, k}} \xi_{m, k},
\end{eqnarray}
where $\xi_{m, k}=\mathbb{E}\left[\left\|\chi_{m, k} y_{m, k}-s_{m, k}\right\|_{2}^{2}\right]$ is the MSE. Substituting (35) into $\xi_{m, k}$, the MSE $\xi_{m, k}$ can be written~as
\begin{flalign}
\!\!\!\!\!\!\!\!\!\!\xi_{m, k}&=\sum_{j=1}^{K}\left|\chi_{m, k} \hat{\mathbf{h}}_{m, k} \mathbf{d}_{m, j}\right|^{2}-2 \Re\left\{\chi_{m, k} \hat{\mathbf{h}}_{m, k} \mathbf{d}_{m, k}\right\}+\left|\chi_{m, k}\right|^{2} \sigma_{m, k}^{2}+1.
\end{flalign}
Then, taking the partial derivatives to (58) with respect to $\chi_{m, k}$ and setting the result to zero, the optimal receive combining filter $\chi_{m, k}^{*}$ at the $k$-th user on the $m$-th subcarrier can be expressed as
\begin{eqnarray}
\chi_{m, k}^{*}=\frac{\hat{\mathbf{h}}_{m, k} \mathbf{d}_{m, k}}{\sum_{j=1}^{K}\left|\hat{\mathbf{h}}_{m, k} \mathbf{d}_{m, j}\right|^{2}+\sigma_{m, k}^{2}}.
\end{eqnarray}
Substituting (59) into (58), the MMSE can be obtained as
\begin{eqnarray}
\xi_{m, k}^{\rm *}=1-\frac{\left|\hat{\mathbf{h}}_{m, k} \mathbf{d}_{m, k}\right|^{2}}{\sum_{j=1}^{K}\left|\hat{\mathbf{h}}_{m, k} \mathbf{d}_{m, j}\right|^{2}+\sigma_{m, k}^{2}},
\end{eqnarray}
which is equal to $\left(1+\gamma_{m, k}\right)^{-1}$, i.e.,
\begin{eqnarray}
\left(1+\gamma_{m, k}\right)^{-1}=\min _{u_{m, k}} \xi_{m, k}.
\end{eqnarray}
Then, the achievable rate of the $k$-th user on the $m$-th subcarrier can be transformed as
\begin{eqnarray}
\log _{2}\left(1+\gamma_{m, k}\right)=\max _{\chi_{m, k}}\left(-\log _{2} \xi_{m, k}\right).
\end{eqnarray}
Based on~\cite{refO1} \cite{refP}, we can obtain
\begin{eqnarray}
\log _{2}\left(1+\gamma_{m, k}\right)=\max _{\chi_{m, k}} \max _{\varpi_{m, k}>0}\left(-\frac{\varpi_{m, k} \xi_{m, k}}{\ln 2}+\log _{2} \varpi_{m, k}+\frac{1}{\ln 2}\right),
\end{eqnarray}
where $\varpi_{m, k}$ is the weight of the data for the $k$-th user on the $m$-th subcarrier and the optimal $\varpi_{m, k}$ is $\varpi_{m, k}^{*}=\frac{1}{\xi_{m, k}}$.
Next, $\mathrm{P2}$ can be transformed to the MSE minimization problem, namely
\begin{subequations}\label{OptA}
\begin{align}
\mathrm{P3}:&\max _{\mathbf{d}_{m, k}} \sum_{k=1}^{K} \sum_{m=1}^{M} \max _{\chi_{m, k}} \max _{\varpi_{m, k}>0}\left(-\frac{\varpi_{m, k} \xi_{m, k}}{\ln 2}+\log _{2} \varpi_{m, k}+\frac{1}{\ln 2}\right)\label{OptA0}\\
{\rm{s.t.}}\;\;&\sum_{k=1}^{K} \sum_{m=1}^{M}\left\|\mathbf{F}_{\rm{A}} \mathbf{F}_{\rm{L}}\mathbf{F}_{\rm{H}} \mathbf{d}_{m, k}\right\|^{2} \leq P_{\text {max}} \label{OptA1}.
\end{align}
\end{subequations}
To address $\mathrm{P3}$, an iterative optimization algorithm is proposed. Based on the obtained $\mathbf{d}_{m, k}^{(i-1)}$ at the $(i-1)$-th iteration, $\chi_{m, k}^{(i)}$ at the $i$-th iteration can be expressed as
\begin{eqnarray}
\chi_{m, k}^{(i)}=\frac{\hat{\mathbf{h}}_{m, k} \mathbf{d}_{m, k}^{(i-1)}}{\sum_{j=1}^{K}\left|\hat{\mathbf{h}}_{m, k} \mathbf{d}_{m, j}^{(i-1)}\right|^{2}+\sigma_{m, k}^{2}}.
\end{eqnarray}
And the optimal $\varpi_{m, k}^{(i)}$ at the $i$-th iteration can be obtained by $\varpi_{m, k}^{(i)}=\frac{1}{\xi_{m, k}^{*(i)}}$
, where
\begin{eqnarray}
\xi_{m, k}^{*(i)}=1-\frac{\left|\hat{\mathbf{h}}_{m, k} \mathbf{d}_{m, k}^{(i-1)}\right|^{2}}{\sum_{j=1}^{K}\left|\hat{\mathbf{h}}_{m, k} \mathbf{d}_{m, j}^{(i-1)}\right|^{2}+\sigma_{m, k}^{2}}.
\end{eqnarray}
Finally, the problem $\mathrm{P3}$ is transformed as
\begin{subequations}\label{OptA}
\begin{align}
\;\;\mathrm{P4}:&\min _{\mathbf{d}_{m, k}^{(i)}} \sum_{k=1}^{K} \sum_{m=1}^{M} \left(\frac{\varpi_{m, k}^{(i)} \xi_{m, k}^{(i)}}{\ln 2}-\log _{2} \varpi_{m, k}^{(i)}-\frac{1}{\ln 2}\right)\label{OptA0}\\
{\rm{s.t.}}\;\;&\sum_{k=1}^{K} \sum_{m=1}^{M}\left\|\mathbf{F}_{\rm{A}} \mathbf{F}_{\rm{L}}\mathbf{F}_{\rm{H}} \mathbf{d}_{m, k}^{(i)}\right\|^{2} \leq P_{\text {max}} \label{OptA1}.
\end{align}
\end{subequations}
We find that $\mathrm{P4}$ can be solved by numerical convex program solvers. Particularly, since the obtained $\mathbf{d}_{m, k}^{(i)}$, $\varpi_{m, k}^{(i)}$, $\chi_{m, k}^{(i)}$ are the optimal solutions of $\mathrm{P4}$ at the $i$-th iteration and the objective function is lower bound and monotonically decreases with iterations. Consequently, it can converge to at least a local optimal solution.
\begin{algorithm}[t]
\caption{ Coordinate Update Algorithm for Optimizing Reflection Coefficients Matrix}
{\bf Input:}
Channels $\mathbf{f}_{r, m, k}$, $\mathbf{G}_{r, m}$, analog beamforming $\mathbf{F}$, digital beamforming vector $\mathbf{d}_{m, k}$, maximum iterations $I_{o}$.\\
{\bf Initialization:}
$\varphi_{n_{\rm ris}}^{(0)}=+1$ for $n_{\rm ris}=1,2,\cdots,R N_{\rm{RIS}}, i=0$.\\
{\bf while} $0 \leq i \textless I_{o}$ {\bf do}\\
\hspace*{0.1in} {\bf for} $n_{\rm ris}=1 : R N_{\rm{RIS}}$ {\bf do}\\
\hspace*{0.2in} $\varphi_{n_{\rm ris}}^{(i)}=-1$;\\
\hspace*{0.2in} $\boldsymbol{\phi}_{1}=\left[\varphi_{1}^{(i)}, \varphi_{2}^{(i)}, \cdots, \varphi_{n_{\rm ris}-1}^{(i)}, \varphi_{n_{\rm ris}}^{(i)},\varphi_{n_{\rm ris}+1}^{(i-1)}\cdots, \varphi_{R N_{\rm{RIS}}}^{(i-1)}\right]^{T}$;\\
\hspace*{0.2in} Transform the vector $\boldsymbol{\phi}_{1}$ to the reflection matrix $\boldsymbol{\Theta}_{1}$;\\
\hspace*{0.2in} $\varphi_{n_{\rm ris}}^{(i)}=+1$;\\
\hspace*{0.2in} $\boldsymbol{\phi}_{2}=\left[\varphi_{1}^{(i)}, \varphi_{2}^{(i)}, \cdots, \varphi_{n_{\rm ris}-1}^{(i)}, \varphi_{n_{\rm ris}}^{(i)},\varphi_{n_{\rm ris}+1}^{(i-1)}\cdots, \varphi_{R N_{\rm{RIS}}}^{(i-1)}\right]^{T}$;\\
\hspace*{0.2in} Transform the vector $\boldsymbol{\phi}_{2}$ to the reflection matrix $\boldsymbol{\Theta}_{2}$;\\
\hspace*{0.2in} $q=\arg \max _{q=1,2}\left\{R_{\rm{sum}}(\boldsymbol{\Theta}_{q})\right\}$;\\
\hspace*{0.2in} $\boldsymbol{\Theta}=\boldsymbol{\Theta}_{q}$;\\
\hspace*{0.1in}{\bf end for}\\
{\bf end while}

{\bf Output:} Reflection coefficients matrix $\mathbf{\Theta}$.
\end{algorithm}
\subsubsection{Optimization of $\mathbf{\Theta}$ with Fixed $\mathbf{d}_{m, k}$}
After obtaining the digital beamforming $\mathbf{d}_{m, k}$, we apply the coordinate update algorithm~\cite{refO0} to obtain the reflection coefficients matrix. Due to $\mathbf{h}_{m, k}=\sum_{r=1}^{R} \mathbf{f}_{r, m, k} \mathbf{\Phi}_{r} \mathbf{G}_{r, m}$, we transform (55) into the following optimization problem:
\begin{subequations}\label{OptA}
\begin{align}
\mathrm{P5}:&\max _{\mathbf{\Theta}} R_{\rm{sum}}\label{OptA0}\\
{\rm{s.t.}}\;\;
&\varphi_{r, m_{x},m_{y}} \in \{e^{j 0}, e^{j{\frac {2 \pi} { 2^{Q}}}}, \cdots, e^{j{\frac {2 \pi} { 2^{Q}}\left(2^{Q}-1\right)}}\}.
\end{align}
\end{subequations}
Considering the discrete nature of the reflection coefficients, the optimal solution for P5 can be obtained by searching all possible vectors. However, even when $Q=1$ bit, such exhaustive search algorithm requires to search $2^{R N_{\rm RIS}}$ candidate vectors, which involves unaffordable complexity for a large  $N_{\rm RIS}$. Therefore, we employ the coordinate update algorithm and it only requires $2 R N_{\rm RIS} I_{o}$ search complexity to obtain a suboptimal solution. Specifically, by fixing any $R N_{\rm RIS}-1$ phase shifts in each iteration,  we alternately optimize each of the $R N_{\rm RIS}$ phase shifts via one-dimensional search over $\mathcal {F}$ in an iterative manner until convergence.

To simplify the expression, we assume $\varphi_{n_{\rm ris}}=\varphi_{r, m_{x},m_{y}}$ and define a vector $\boldsymbol{\phi}\in \mathbb{C}^{R N_{\rm{RIS}} \times  1}=\left[\varphi_{1}, \varphi_{2}, \cdots, \varphi_{n_{\rm ris}-1}, \varphi_{n_{\rm ris}}, \varphi_{n_{\rm ris}+1}, \cdots, \varphi_{R N_{\rm{RIS}}}\right]^{T}$, $n_{ris}=1,2,\cdots,R N_{\rm{RIS}}$. Without loss of generality, we set $Q=1$ bit. The optimization problem is transformed as
\begin{subequations}\label{OptA}
\begin{align}
\mathrm{P6}:&\max _{\boldsymbol{\phi}} R_{\rm{sum}}\label{OptA0}\\
{\rm{s.t.}}\;\;
&\varphi_{n_{\rm ris}} \in \{1,-1\}.
\end{align}
\end{subequations}
Based on the coordinate update algorithm, the optimal phase shift of the $n_{\rm ris}$-th element is given by
\begin{eqnarray}
\varphi_{n_{\rm ris}}=\arg \max _{\varphi_{n_{\rm ris}}\in {\{1, -1\}}}\left\{R_{\rm{sum}}(e^{j\varphi_{n_{\rm ris}}})\right\}.
\end{eqnarray}
We summarize the procedure for solving reflection coefficients in $\mathbf{Algorithm}$ $\mathbf{1}$.
Note that problem P1 is the original problem formulation. P2-P6 are the intermediate problems obtained by transformation and simplification in the solving process. The digital beamforming vector and reflection coefficients matrix can be obtained by solving P4 and P6, respectively.

So far, the analog beamforming $\mathbf{F}$, the digital beamforming $\mathbf{d}_{m, k}$ and reflection coefficients matrix $\mathbf{\Theta}$ are all obtained. In the process of solving P1, based on the RISs' locations, we first design the analog beamforming, including PSs' phase shifts and time delays of the double-layer TTD network. Then, we propose an alternatively optimization algorithm to obtain the digital beamforming and reflection coefficients. Specifically, given fixed reflection coefficients, the MMSE technique is applied to obtain the digital beamforming. Next, we employ the coordinate update algorithm to slove the reflection coefficients. The above procedure is repeated until convergence. The details of the proposed optimization framework are summarized in $\mathbf{Algorithm}$~$\mathbf{2}$, which can obtain a sub-optimal solution to balance the performance and computation complexity.
\begin{algorithm}[t]
\caption{The Proposed Algorithm for Solving $\mathrm{P1}$}
{\bf Input:}
Channels $\mathbf{f}_{r, m, k}$, $\mathbf{G}_{r, m}$.\\
{\bf Initialization:}
Digital beamforming vector $\mathbf{d}_{m, k}^{(0)}$ and reflection coefficients matrix $\mathbf{\Theta}^{(0)}$.

Calculate the analog beamforming matrix $\mathbf{F}$ according to (48), (50) and (54).\\
{\bf while} $0 \leq i \textless I_{\rm {max}}$ {\bf do}\\
\hspace*{0.1in} Obtain digital beamforming vector $\mathbf{d}_{m, k}$ by solving P4;\\
\hspace*{0.1in} Obtain reflection coefficients matrix $\mathbf{\Theta}$ by solving P6;\\
{\bf end while}

{\bf Output:}
Analog beamforming matrix $\mathbf{F}$, digital \hspace*{0.02in} beamforming vector $\mathbf{d}_{m, k}$, reflection coefficients matrix $\mathbf{\Theta}$.
\end{algorithm}
\subsection{Computational Complexity}
In this subsection, we analyze the computational complexity of the proposed algorithm, which is mainly induced by MMSE algorithm and coordinate update algorithm. Specifically, to obtain the digital beamforming vector $\mathbf{d}_{m, k}$, the computational complexity is $\mathcal{O}\left(I_d M N^{2}\right)$, where $I_{d}$ is the required number of iterations of the MMSE algorithm. The computational complexity is $\mathcal{O}\left(I_{o} 2^Q K N R N_{\rm{RIS}}\right)$ for solving the reflection coefficients matrix $\boldsymbol{\Theta}$, where $I_{o}$ is the required number of iterations of the coordinate update algorithm. The overall computational complexity of the proposed $\bf Algorithm \hspace*{0.02in} 1$ is $\mathcal{O}\left(I_{\rm{max}}(I_d M N^{2}+I_{o} 2^QK N R N_{\rm{RIS}}) \right)$, where $I_{\rm {max}}$ is the required iteration number of the outer iteration.
\section{Numerical Results}
In this section, simulation results are presented to evaluate the performance of the proposed schemes.
We assume that the BS is located at (50m, 0m, 3m) and $K=4$ users are randomly distributed in a circle centered at (0, 85m, 0) with radius of 1m. In addition, we deploy $R=4$ distributed RISs and their locations are (0, 80m, 6m), (0, 80m, 8m), (0, 85m, 6m) and (0, 85m, 8m), respectively. Since THz communication mainly relies on the LoS path, we set $L_{1}=L_{2}= 1$~\cite{refG3}.  The other default simulation parameters are listed in Table~I.
\begin{table}[htb]
\begin{center}
\caption{System parameters}
\label{table:1}
\begin{tabular}{|c|c|c|}
\hline   \textbf{Parameters} & \textbf{Value} \\
\hline   Number of antennas &  $N=128$ \\
\hline   Central frequency & $f_{c}=300$ GHz  \\
\hline   Bandwidth & $B=30$ GHz  \\
\hline   Number of subcarriers & $M=8$  \\
\hline   Bit of the TTD at the first layer &  $P_{\rm{H}}=8$ \\
\hline   Bit of the TTD at the second layer &  $P_{\rm{L}}=4$ \\
\hline   Number of RISs &  $R=4$ \\
\hline   Number of RIS elements &  $N_{\rm{RIS}}=16$ \\
\hline   Number of users &  $K=4$ \\
\hline   Number of RF chains & $N_{\rm{RF}}=4$  \\
\hline   Maximum transmit power & $P_{\rm{max}}=10$ dBm  \\
\hline   Noise power & $\sigma_{m,k}^{2} = -85$ dBm  \\
\hline
\end{tabular}
\end{center}
\end{table}

\begin{figure}[htbp]
	\centering
		\label{complexity} 
		\includegraphics[width=9.5cm,height=6.5cm]{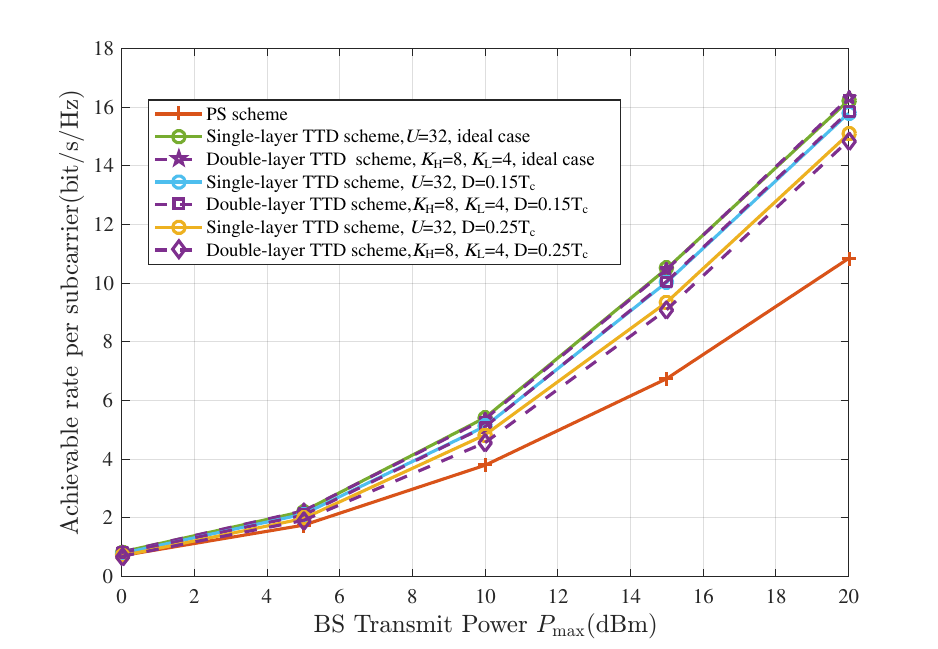}
	\caption{Achievable rate versus the BS transmit power under different time delay steps.}
\end{figure}
Fig. 9 plots the achievable rate versus maximum transmit power $P_{\rm {max}}$ with different time delay steps. Here, we first consider the infinite-resolution phase shift at PSs. In addition, we mainly focus on the performance of the proposed double-layer TTD scheme at the BS, and thus the RIS is not considered in this simulation. For the single- and double-layer TTD schemes, we set $U=32$, $K_{\rm H}=8$, $K_{\rm L}=4$, $P_s=8$, $P_{\rm{H}}=8$, $P_{\rm{L}}=4$. The time delay step $D$ includes ideal case (continuous), $0.15T_{c}$, and $0.25T_{c}$. One can observe that the achievable rate of the single- and double-layer  TTD schemes increases with the time delay step decreases. The reason is that the accuracy of the TTD becomes poor with a large time delay step, which results in the beam misalignment. Moreover, we find that  the double-layer TTD scheme can almost obtain the same performance with the single-layer TTD scheme, but the number of large-range delay TTDs and the total number of bits can be effectively reduced. In addition, it is obvious that the PS scheme is the worst due to the serious beam split effect.

Fig. 10 (a) and Fig. 10 (b) show the convergence performance of the  proposed inner iterative algorithm for solving the digital beamforming and reflection matrix, respectively,  i.e., lines 5 and lines 6 in Algorithm 2. We set $K_{\rm H}=8$, $K_{\rm L}=4$ in the double-layer TTD scheme. Besides, we assume $D=0.15T_{c}$, $P_{\rm{H}}=8$, $P_{\rm{L}}=4$, $P_{\rm{s}}=8$, $Q=1$ and infinite-resolution phase shift at PSs. The legend ``$n$-th  iteration" in Fig. 10 stands for the outer iteration number. One can observe that the inner iterative algorithm tends to converge after five iterations for each outer iteration. In addition, it can be found that the gap is small between the second and third iterations, but large between the first and second iterations. This means that outer iterative loop (i.e., Algorithm 2) also converges rapidly.
\begin{figure}[htbp]
	\centering
		\label{complexity} 
        \subfigure[]
		{\includegraphics[width=8cm,height=5cm]{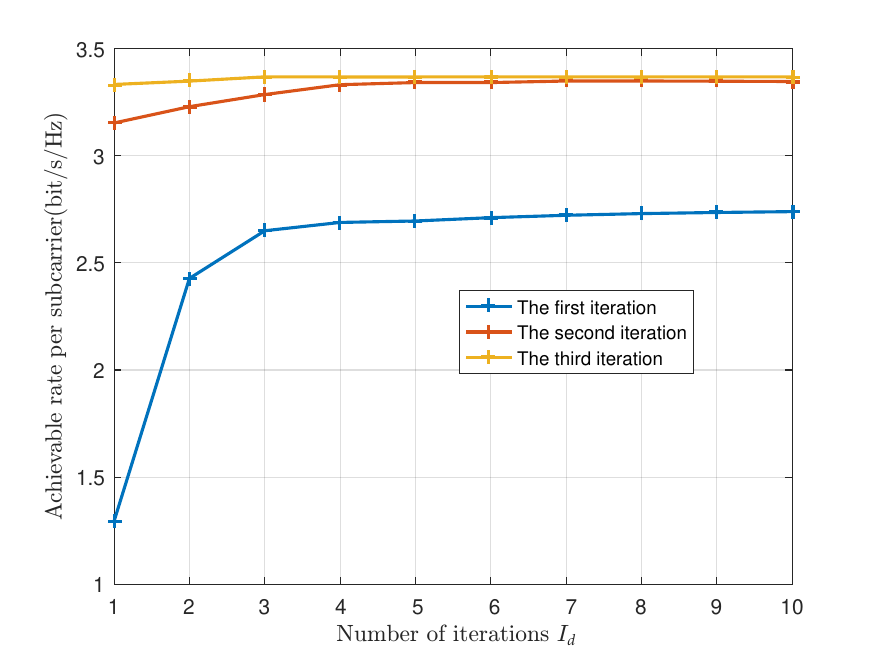}}
	\centering
		\label{complexity} 
        \subfigure[]
		{\includegraphics[width=8cm,height=5cm]{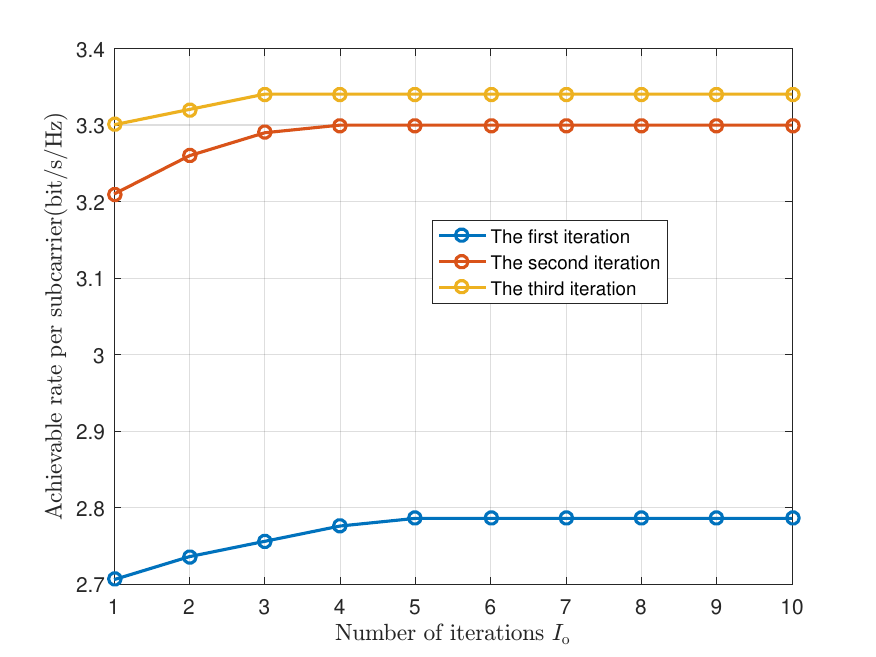}}
	\caption{Achievable rate versus iteration for solving (a) the digital beamforming, (b) the reflection coefficients matrix.}
\end{figure}

\begin{figure}[htbp]
	\centering
		\label{complexity} 
		\includegraphics[width=9.5cm,height=6.7cm]{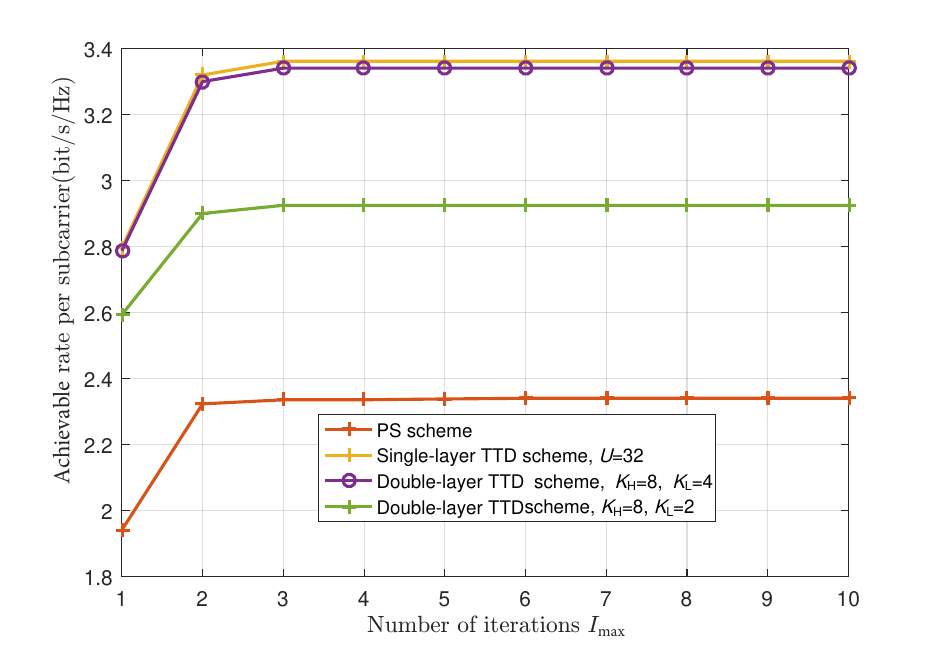}
	\caption{Achievable rate versus the iteration.}
\end{figure}

\begin{table*}[htb]
\begin{center}
\caption{Hardware complexity and performance comparison}
\label{table:1}
\begin{tabular}{|c|c|c|c|c|c|}
\hline    & & \textbf{Number of large-} & & &\textbf{Achievable} \\
\textbf{Architecture }& \textbf{Delay range}&\textbf{range delay TTDs}&\textbf{Total number of TTDs }&\textbf{Total number of bits }&\textbf{rate}\\
\hline Single-layer TTD scheme &  &   &  &  &\\
 ($U=32$) &$\tau_{u} \in [0,62 T_c]$& $N_{\rm RF} U=128$ & $N_{\rm RF} U=128$  & $N_{\rm RF} U P_s=1024$ & $3.36$ bit/s/Hz \\
\hline Double-layer TTD scheme &$\tau_{n_{\rm rf}, k_h} \in [0, 56 T_c]$  &  &  & &\\
($K_{\rm H}=8$, $K_{\rm L}=4$) &$\tau_{n_{\rm rf}, k_h, k_l} \in [0, 6 T_c]$& $N_{\rm RF}K_{\rm H}=32$ & $N_{\rm RF}(K_{\rm H} + K_{\rm H} K_{\rm L})=160$  &  $N_{\rm RF}(K_{\rm H} P_{\rm H} + K_{\rm H} K_{\rm L} P_{\rm L})=768$&  $3.34$ bit/s/Hz\\
\hline Double-layer TTD scheme &$\tau_{n_{\rm rf}, k_h} \in [0, 56 T_c]$ &   &  & &\\
($K_{\rm H}=8$, $K_{\rm L}=2$) &$\tau_{n_{\rm rf}, k_h, k_l} \in [0, 4 T_c]$& $N_{\rm RF}K_{\rm H}=32$& $N_{\rm RF}(K_{\rm H} + K_{\rm H} K_{\rm L})=96$  &  $N_{\rm RF}(K_{\rm H} P_{\rm H} + K_{\rm H} K_{\rm L} P_{\rm L})=512$&  $2.93$ bit/s/Hz\\
\hline
\end{tabular}
\end{center}
\end{table*}

Fig. 11 shows the achievable rate versus outer iterations $I_{\rm{max}}$ under different schemes. We set $U=32$ in the single-layer TTD scheme serves as the performance upper bound. The classical PS scheme is adopted as the baseline. And  we set $K_{\rm H}=8$, $K_{\rm L}=4$ and $K_{\rm H}=8$, $K_{\rm L}=2$ two cases in the double-layer TTD scheme. Meanwhile, we assume $D=0.15T_{c}$, $P_{\rm{H}}=8$, $P_{\rm{L}}=4$, $P_{\rm{s}}=8$, $Q=1$ and infinite-resolution phase shift at PSs. One can observe that the achievable rate tends to convergence after 3 iterations, which proves the effectiveness of the proposed algorithm. Besides, when the number of large-range delay TTDs $K_{\rm H}$ is fixed, the achievable rate increases with the increased number of small-range delay TTDs. Specifically, the proposed double-layer TTD scheme with $K_{\rm L}=2$ small-range delay TTDs can significantly  enhance the performance. And when $K_{\rm L}=4$, the double-layer TTD scheme is almost the same with that of the single-layer TTD scheme, which proves that the double-layer TTD scheme is another hardware-efficient approach to solve beam split effect. The reason is that the proposed small-range delay TTDs only need to compensate the propagation delay across the small subarray aperture.

Next, we compare the hardware complexity and the achievable rate under single- and double-layer TTD schemes as shown in TABLE II. One can observe that the number of large-range delay TTDs reduces from 128 under the single-layer TTD scheme to 32 under the proposed double-layer TTD scheme, which is down by $75\%$. In addition, the total number of bits is down by $25\%$ for $K_{\rm H}=8$ and $K_{\rm L}=4$, but the achievable rate is only down by $0.6\%$ for the proposed scheme. When $K_{\rm H}=8$ and $K_{\rm L}=2$, there is a more obvious advantage. For example, the total number of bits is down by $50\%$ while  the achievable rate is a little decrease. Furthermore, although an additional small-range delay TTD network is introduced, its required delay range is much smaller compared to large-range delay TTD network. Therefore, the proposed scheme can effectively reduce the hardware cost by sacrificing a little achievable rate.

In Fig.~12, we present the achievable rate versus the BS transmit power under different schemes. We set $D=0.15T_{c}$, $P_{\rm{H}}=8$, $P_{\rm{L}}=4$, $P_{\rm{s}}=8$, $Q=1$, and infinite-resolution phase shifter at PSs. One can observe that the achievable rate increases with the transmit power under all schemes. In addition, the double-layer TTD scheme with $K_{\rm H}=8$ and $K_{\rm L}=4$ can almost obtain the same performance with the single-layer TTD scheme, while the achievable rate of the double-layer TTD scheme with $K_{\rm H}=8$ and $K_{\rm L}=2$ has a little decrease. However, according to TABLE II, comparison with single-layer TTD scheme, the decreasing percentage of the hardware complexity is much larger than that of the achievable rate for the proposed double-layer TTD scheme.
\begin{figure}[htbp]
	\centering
		\label{complexity} 
		\includegraphics[width=9.5cm,height=6.7cm]{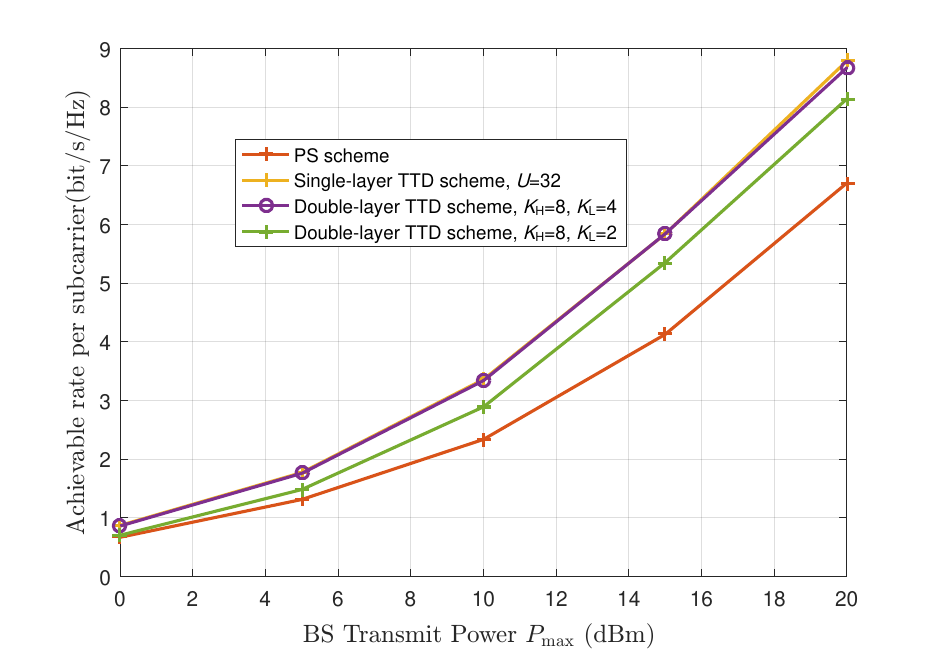}
	\caption{Achievable rate versus the BS transmit power under different schemes.}
\end{figure}

Fig. 13 shows the achievable rate versus BS transmit power with different number of RIS elements $R_{\rm{total}}$. Here, we set $D=0.15T_{c}$, $P_{\rm{H}}=8$, $P_{\rm{L}}=4$, $P_{\rm{s}}=8$, $Q=1$, and infinite-resolution phase shift at PSs. It is obvious that more RIS elements can obtain a higher rate under the same conditions. Besides, it demonstrates that the proposed framework can be applied to any number of RIS elements.
Fig.~14 plots the achievable rate versus iterations $I_{\rm{max}}$, where we set $D=0.15T_{c}$, $P_{\rm{H}}=8$, $P_{\rm{L}}=4$, $P_{\rm{s}}=8$, $Q=1$, $b=1$. One can observe that the achievable rate also tends to convergence after 3 iterations, which proves the effectiveness of the proposed method. In addition, it can be observed that the achievable rate is a little lower than that of the infinite-resolution phase shift at PSs presented in Fig. 11. However, it is acceptable for a little degradation in achievable rate to obtain a large reduction in hardware complexity and cost.
\begin{figure}[htbp]
	\centering
		\label{complexity} 
		\includegraphics[width=9.5cm,height=6.5cm]{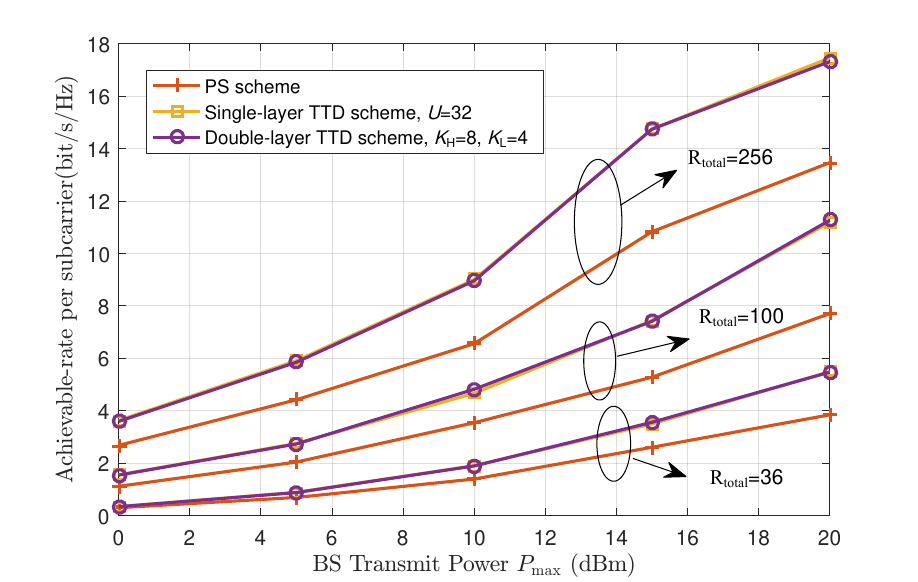}
	\caption{Achievable rate versus the BS transmit power with different RIS elements.}
\end{figure}
\begin{figure}[htbp]
	\centering
		\label{complexity} 
		\includegraphics[width=9.5cm,height=6.7cm]{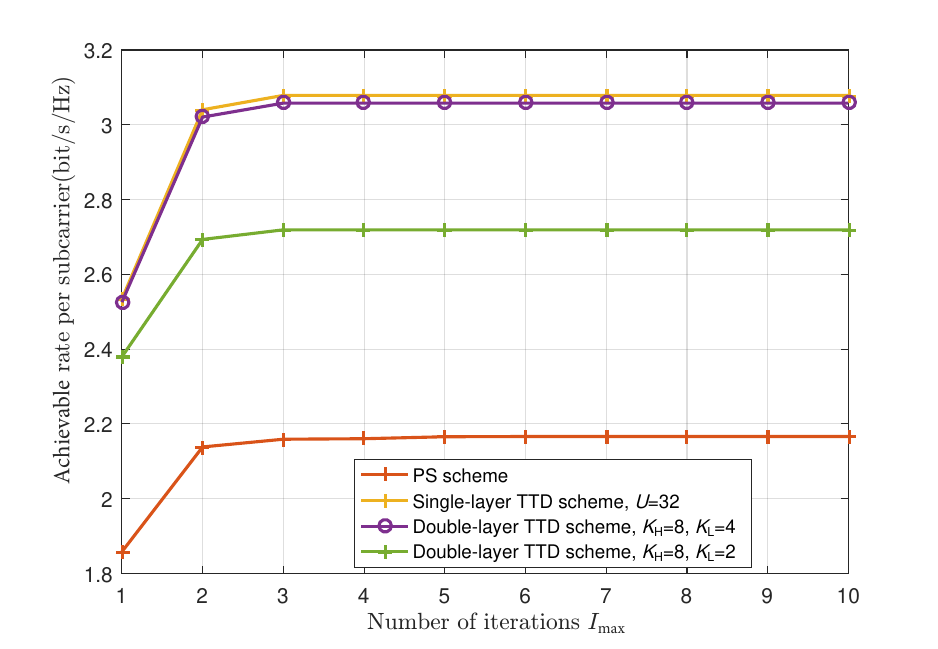}
	\caption{Achievable rate versus the iteration with $b=1$.}
\end{figure}

Furthermore, we analyze the robustness of the proposed joint wideband beamforming against the impacts of imperfect CSI.
We consider CSI uncertainties that can be modeled as
\begin{eqnarray}
\widetilde h=h+e,
\end{eqnarray}
where $\widetilde h$ and $h$ denote estimated and real channel, respectively, $e$ represents the independent estimation error following complex Gaussian distribution with zero mean, i.e. $e \sim \mathcal{C} \mathcal{N}\left(0, \sigma_{e}^{2}\right)$. We assume that the variance $\sigma_{e}^{2}$ , i.e. the error power, satisfies $\sigma_e^2 \triangleq \delta|h|^2$ where $\delta$ denotes the ratio of the error power $\sigma_{e}^{2}$ to the channel gain $|h|^2$, which  characterizes the level of CSI~error.

We assume $D=0.15T_{c}$, $P_{\rm{H}}=8$, $P_{\rm{L}}=4$, $P_{\rm{s}}=8$, $Q=1$ and $b=1$. Then, the achievable rate per subcarrier versus the CSI error parameter $\delta$ is shown in Fig. 15. One can observe that the performance loss grows with the increasing of $\delta$. The reason is that the accuracy of  the estimation angles becomes poor with a large error, which  results in the beam misalignment. For example, for the ``Double-layer TTD scheme, $K_{\rm H}=8$, $K_{\rm L}=4$", compared with the perfect CSI without error (i.e. $\delta = 0$), the system performance suffers a loss of $6\%$ when the error power $\delta = 0.1$, and a loss of $30\%$ when $\delta = 0.3$. Besides, the proposed double-layer TTD scheme always outperforms the frequency-independent PS scheme without TTD  and close to the single-layer TTD scheme at any CSI estimation error, which validates the effectiveness and robustness of our proposed joint wideband beamforming to deal with beam split.

\begin{figure}[htbp]
	\centering
		\label{complexity} 
		\includegraphics[width=9.5cm,height=6.5cm]{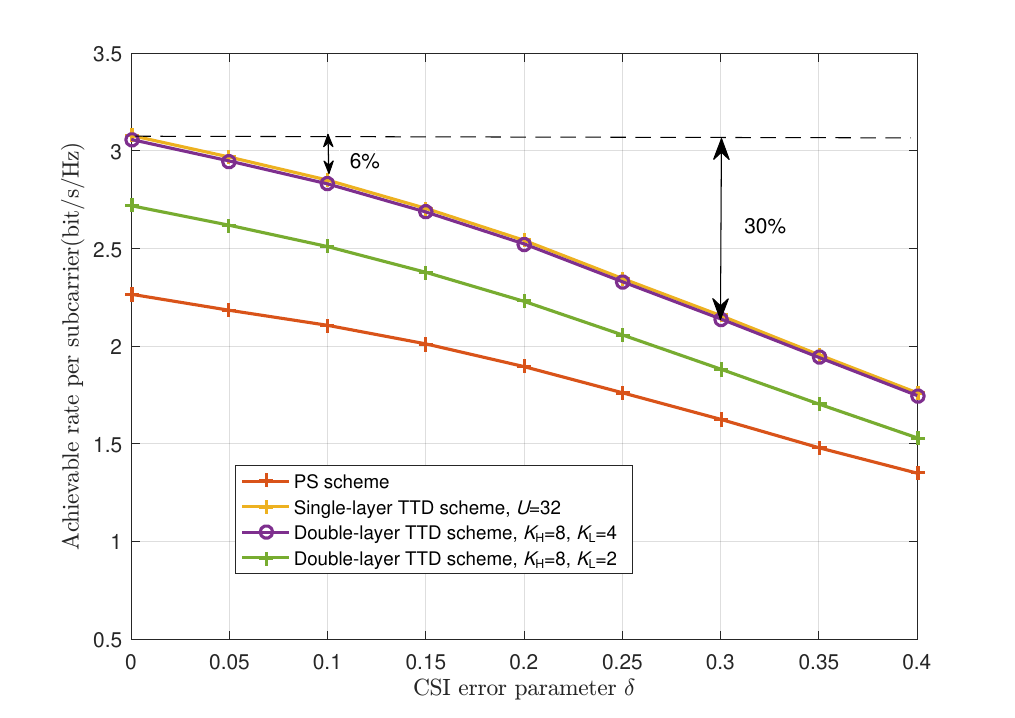}
	\caption{Average achievable rate per subcarrier versus $\delta$.}
\end{figure}

\section{Conclusions}
In this paper, we proposed a double-layer TTD scheme  with low hardware cost to overcome the beam split of the BS and solve the maximum delay compensation problem observed in the traditional single-layer TTD scheme. We first analyzed the phase compensation error and normalized array gain under the double-layer TTD scheme. Then, based on the proposed scheme, we investigated the beamforming optimization problem for the multiple distributed RISs-aided THz communications and formulated a achievable rate maximization problem via jointly optimizing the hybrid analog/digital beamforming, time delays of the double-layer TTD network and reflection coefficients of the RISs. Theoretical analysis and simulation results demonstrated that the double-layer TTD scheme can almost obtain the same performance with the single-layer TTD scheme, while the overall hardware cost is effectively decreased. In our future work, to further reduce the time delay range of the TTD at the BS, we will extend the proposed double-layer TTD network to the sub-connected hybrid beamforming architecture. Besides, to solve the beam split at the RIS,  we will introduce TTDs to RIS elements and research how to reduce the beam split effect.

\end{document}